\documentclass[aps,pre,twocolumn,superscriptaddress]{revtex4}
\usepackage[utf8]{inputenc}
\usepackage{amsmath}
\usepackage{amssymb}
\usepackage{mathtools}
\usepackage{graphicx}
\usepackage{cases}
\usepackage{color}
\usepackage{multirow} % Required for multirows
\usepackage{hhline}

\usepackage[pdfencoding=auto,psdextra]{hyperref}

\DeclareMathOperator{\tr}{tr}
\DeclareMathOperator{\sgn}{sgn}

\renewcommand{\Re}{\mathop\mathrm{Re}\nolimits}

\begin{document}

\title{Mesoscopic supercurrent fluctuations in diffusive magnetic Josephson junctions}

\author{P.~A.\ Ioselevich}
\affiliation{National Research University Higher School of Economics, 101000 Moscow, Russia}
\affiliation{L.~D.\ Landau Institute for Theoretical Physics RAS, 142432 Chernogolovka, Russia}

\author{P.~M.\ Ostrovsky}
\affiliation{Max Planck Institute for Solid State Research, Heisenbergstr.\ 1,
70569 Stuttgart, Germany}
\affiliation{L.~D.\ Landau Institute for Theoretical Physics RAS, 142432 Chernogolovka, Russia}

\author{Ya.~V.\ Fominov}
\affiliation{L.~D.\ Landau Institute for Theoretical Physics RAS, 142432 Chernogolovka, Russia}
\affiliation{\color{black} Moscow Institute of Physics and Technology, 141700 Dolgoprudny, Russia}
\affiliation{National Research University Higher School of Economics, 101000 Moscow, Russia}

\begin{abstract}
We study the supercurrent in quasi-one-dimensional Josephson junctions with a weak link involving magnetism, either via magnetic impurities or via ferromagnetism. In the case of weak links longer than {\color{black}the magnetic pair-breaking} length, the Josephson effect is dominated by mesoscopic fluctuations. We establish the supercurrent-phase dependence $I(\varphi)$ along with statistics of its sample-dependent properties in junctions with transparent contacts between leads and link. High transparency gives rise to the inverse proximity effect, while the direct proximity effect is suppressed by magnetism in the link. We find that all harmonics are present in $I(\varphi)$. Each harmonic has its own sample-dependent amplitude and phase shift with no correlation between different harmonics. 
{\color{black} Depending on the type of magnetic weak link, the system can realize a $\varphi_0$ or $\varphi$ junction in the fluctuational regime.}
Full supercurrent statistics is obtained at arbitrary relation between temperature, superconducting gap, and the Thouless energy of the weak link.
\end{abstract}

\maketitle

\tableofcontents

\section{Introduction}

The Josephson effect allows Cooper pairs to flow between two superconductors (S) connected by a weak link, which does not need to be superconducting itself. This produces a stationary macroscopic supercurrent between the superconductors. In particular, the weak link can be represented by a normal metal (N).
\textit{s}-wave superconductivity and the Josephson current are robust against potential impurity scattering \cite{AG1958-59,Anderson1959}, and the phase-dependent supercurrent $I(\varphi)$ persists even in the diffusive limit, which we consider in this work ($\varphi$ is the phase difference between the superconductors).
Details of the Josephson effect in an SNS junction with a weak link of length $L$ then depend on the relation between essential energy scales: temperature $T$, pair potential $\Delta$, and the Thouless energy $E_\mathrm{Th} =  D/L^2$ (the inverse diffusion time through the weak link with diffusion constant $D$) \cite{GolubovReview2004}.

In the presence of disorder, the Josephson current $I(\varphi)$ depends on specific impurity configuration.
In addition to the disorder-averaged supercurrent $\langle I(\varphi)\rangle$, there are also sample-dependent mesoscopic fluctuations $\delta I(\varphi)\equiv I(\varphi)-\langle I(\varphi)\rangle$ (for comprehensive discussion of this effect in SNS junctions, see Ref.\ \cite{SkvortsovHouzet} and references therein). Characteristics of the fluctuational current, in particular, its scaling with the junction cross-section (i.e., the number $N$ of conducting channels) depend on the cross-section itself.
We focus on the case of quasi-one-dimensional (q1D) wires, in which case $N$ can be large but the cross-section is still relatively small so that soft transverse diffusive modes are irrelevant.
While the average current is proportional to $N$, the fluctuational current $\delta I(\varphi)$ in q1D junctions does not scale with $N$ (this underlines that mesoscopic fluctuations of the Josephson current are of similar nature with universal conductance fluctuations \cite{AltshulerUCF1985,LeeStoneUCF1985}). In the case of $N\gg 1$, the fluctuations are therefore small compared to the average current.

However, if Cooper pairs are destroyed inside the weak link because of broken time-reversal symmetry (TRS), the average supercurrent is strongly suppressed, $\langle I(\varphi) \rangle \propto \exp(-L/l_*)$, where $l_*$ is the pair-breaking length.
At the same time, certain types of fluctuations survive in this case.
TRS is naturally broken by magnetism-related physics. Typical examples are
(a)~magnetic impurities, (b)~ferromagnetism, and (c)~external magnetic field.

As a result, in magnetic junctions, the average supercurrent can become smaller than its fluctuations: $\langle I\rangle\ll \delta I$ or, equivalently, $\langle I\rangle\ll \sqrt{\langle I^2\rangle}$. This means that the current is strongly sample-dependent and the averaged current is vanishingly small, while the typical current (determined by mesoscopic fluctuations) is much larger.

The fluctuational current in magnetic Josephson junctions has been calculated analytically in the tunneling limit (low transparency of interfaces between the superconducting and magnetic parts of the junction) in Ref.\ \cite{Zyuzin2003} and illustrated numerically in Ref.\ \cite{Asano2007}. In Refs.\ \cite{Melin2005,we2017}, the fluctuational current in the tunneling limit was studied with special focus on interferometer-type geometry of magnetic links, allowing processes with splitting of Cooper pairs between magnetic arms.

At the same time, the fluctuational supercurrent should be largest in the opposite limit of junctions with transparent interfaces. This limit is therefore the most promising from the point of view of experimental observability.
The fluctuational current in magnetic junctions in the transparent limit was studied in Ref.\ \cite{Altshuler1987} both in the q1D and in the wide-junction regime (the latter case requires taking transverse diffusion modes into account, which results in $\sqrt{N}$ scaling of the fluctuational current); the wide-junction regime was also considered in Ref.\ \cite{Melnikov2016}.
In the transparent limit, in addition to penetration of superconductivity into the weak link (proximity effect), the inverse proximity effect arises, meaning that the magnetic region can suppress superconducting correlations (described by the anomalous Green function) in the superconducting parts near the interfaces \cite{BuzdinReview2005,BVE}. The consideration of Ref.\ \cite{Altshuler1987} assumed the case of strong superconductors with effectively infinite pair potential $\Delta$ (more accurately, $\Delta\gg T, E_\mathrm{Th}$) and the inverse proximity effect was neglected.

In this paper, we investigate the fluctuational Josephson current in q1D junctions in the limit of transparent interfaces taking the inverse proximity effect into account at arbitrary relation between $\Delta$, $T$, and $E_\mathrm{Th}$.
We consider the situation where the inverse proximity effect leads to spatial variations of average Green functions in the superconducting banks of the junction, while the pair potential $\Delta$ remains uniform in the superconductors. Thus, the overall spatial dependence of $\Delta$ across the system is steplike, see Fig.~\ref{fig1}. We calculate the current-current correlator $\langle I(\varphi_1) I(\varphi_2)\rangle$, from which we extract the current-phase relation and statistical properties of the sample-dependent Josephson current.
In contrast to the sinusoidal current-phase relation in the tunneling limit \cite{Zyuzin2003,Melin2005,we2017}, the fluctuational current in the limit of transparent interfaces can also contain higher harmonics.

Hybrid proximity structures involving superconducting and magnetic parts currently attract great attention
 as a platform for realization of Majorana fermions (see, e.g., Ref.\ \cite{LutchynReview2018} for a recent review).
Such systems are engineered from thin wires with superconductivity induced
by a substrate (or cover). In this case, the effective pair potential $\Delta$ is imposed by the substrate and has a steplike spatial profile along the wire, as in Fig.~\ref{fig1}, while the interfaces between the segments of the wire are transparent by design.
These are exactly the conditions we assume.

In Sec.~\ref{secMethod}, we introduce the sigma model used to study the Josephson transport in the system.
In Sec.~\ref{secSaddle}, we describe the saddle-point solution of the sigma model, which corresponds to the quasiclassical description of the system.
In Sec.~\ref{secFluct}, we develop the model to include fluctuations around the saddle-point.
In Sec.~\ref{secCorr}, we calculate the current-current correlator $\langle I(\varphi_1)I(\varphi_2)\rangle$ which we then use in Sec.~\ref{secStat} to extract the current-phase relations and statistical properties of the sample-dependent supercurrent for various cases and limits.
In Sec.~\ref{secF} we discuss the differences between systems with different magnetic links and present results for the ferromagnetic case.
In Sec.~\ref{secDisc}, we discuss the results.
Finally, in Sec.~\ref{secConclusion}, we present our conclusion. Throughout the paper, we employ the units with $k_B =\hbar=1$.

\section{Method}\label{secMethod}

\begin{figure}[t]
\centering
\hspace*{-3pt}\includegraphics[width=0.485\textwidth]{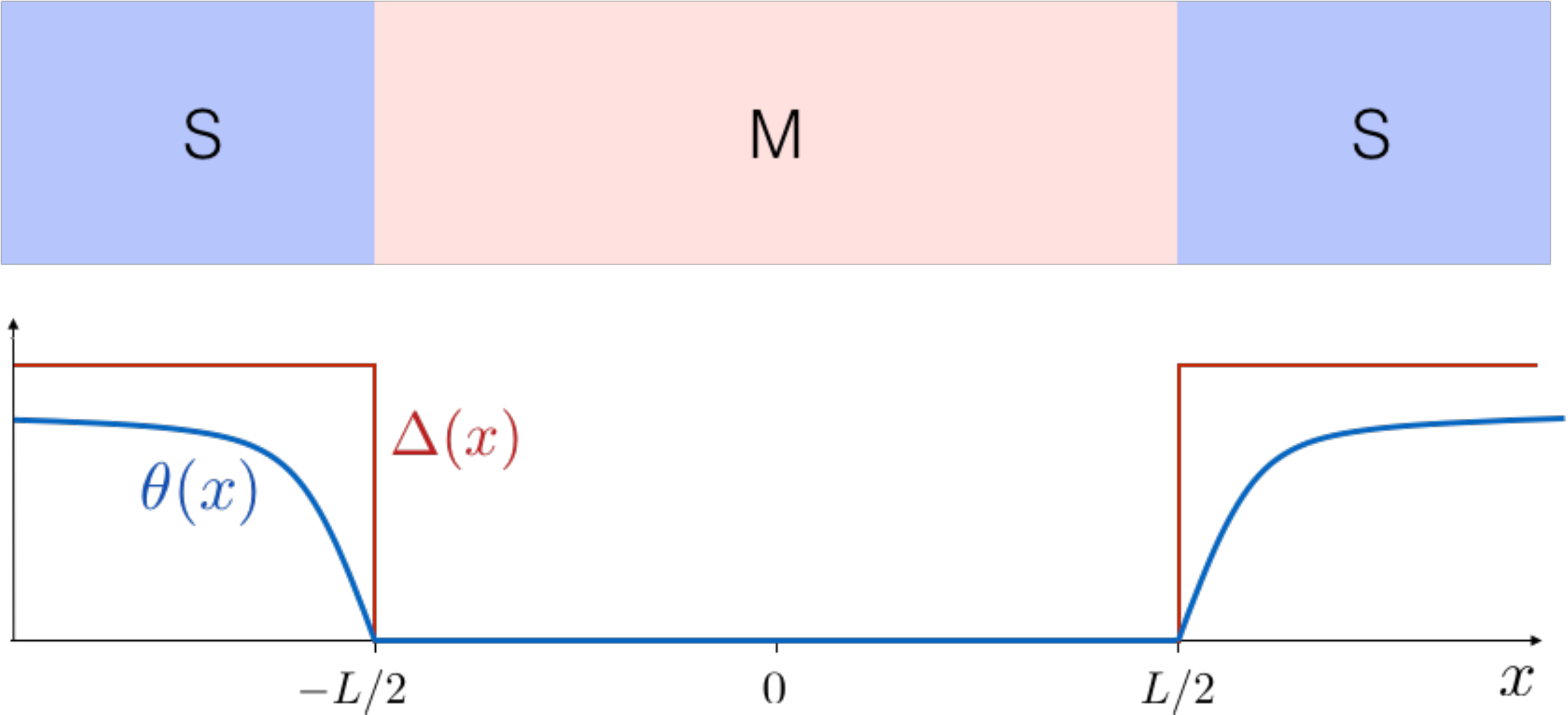}
\caption{
Top: q1D junction. Superconducting leads (S) are connected by a magnetic region made of either a metal with magnetic impurities (M), or a metal in an external magnetic field, or a ferromagnet. The junction is long compared to the characteristic {\color{black} pair-breaking} length $l_*$. Bottom: solution $\theta(x)$ of the Usadel equation (i.e., the saddle-point configuration of the sigma model).
}
\label{fig1}
\end{figure}

We consider a strongly diffusive system, where the mean free path $l$ is the shortest scale (except the Fermi wave length). In this so-called dirty limit, $\langle I(\varphi)\rangle$ can be readily obtained from the Usadel equation \cite{Usadel1970}, which operates in terms of Green's functions averaged over disorder. To access fluctuations of the current, we must calculate a higher-order correlator, $\langle I(\varphi_1) I(\varphi_2)\rangle$. Using the general relation to free energy, $I(\varphi)=2e\partial F/\partial\varphi$, we write the correlator as
\begin{equation}
\langle I(\varphi_1)I(\varphi_2)\rangle = 4e^2\frac{\partial^2 \langle F(\varphi_1)F(\varphi_2)\rangle }{\partial\varphi_1\partial\varphi_2}.
\end{equation}
The average $\langle F(\varphi_1)F(\varphi_2)\rangle$ is a second-order Green's function correlator and lies beyond the scope of the Usadel equation. To calculate it, we employ the nonlinear sigma model \cite{Efetov}. It is a powerful approach for describing systems in the dirty limit. We use the replica sigma model \cite{Wegnersigma}, which formulates the above correlator as a functional integral
\begin{equation}
\langle I(\varphi_1)I(\varphi_2)\rangle = 4e^2T^2\sum\limits_{\omega_1,\omega_2}\frac{\partial^2 }{\partial\varphi_1\partial\varphi_2}\lim\limits_{n_{1,2}\to0}\frac{\int e^{-S[Q]}DQ}{n_1n_2}. \label{IQ}
\end{equation}
The currents $I(\varphi_1)$ and $I(\varphi_2)$ are the results of summation over the Matsubara frequencies $\omega_1$ and $\omega_2$, respectively. The field $Q(x)$ is a matrix in replica space containing $n_1$ replicas corresponding to $I(\varphi_1)$ and another $n_2$ replicas corresponding to $I(\varphi_2)$.

We model our magnetic junction as a q1D wire with superconducting and magnetic parameters varying in space along $x$. The corresponding sigma-model action is
\begin{equation} \label{act1}
S[Q] = \frac{\pi\nu}{8} \int dx \tr \left\{
D (\nabla Q)^2
- 4 \left(\hat\omega\Lambda + \check\Delta(x)\right)  Q\right\},
\end{equation}
with $\Lambda \equiv \sigma_3\tau_3$, supplemented by the constraint
\begin{equation} \label{Qtau3}
[Q,\tau_3]=0
\end{equation}
in the magnetic region. $Q(x)$ is a matrix in Nambu ($\tau_i$), particle-hole ($\sigma_i$), and replica space obeying $Q^2=1$. The constraint Eq.\ \eqref{Qtau3} means that off-diagonal components of $Q$ in Nambu space are suppressed. This is how we implement the effects of magnetism in the link and corresponds to neglecting the proximity effect, i.e., the penetration of superconductivity into the magnetic region. This is justified by our assumption that the proximity effect decays on a scale $l_*$ much shorter than the link length $L$, so that its contribution to the Josephson effect is exponentially small, $\propto e^{-L/l_*}$. A detailed justification of Eq.\ \eqref{Qtau3} is given in Appendix~\ref{appModel} together with a derivation of Eq.\ \eqref{act1}. Throughout this paper we use the hat, as in Matsubara frequency $\hat\omega$, to denote variables that depend on replica (and are thus a diagonal matrix in replica space) such as the Matsubara frequency $\hat\omega$.

The particle-hole space is usually introduced in a sigma model to take into account the Bogoliubov--de Gennes symmetry of the problem \cite{Altlandsigma}. At the same time, it doubles as the space of positive and negative frequencies. Thus, to avoid double-counting, the sum in Eq.\ \eqref{IQ} is taken only over positive $\omega_{1,2}$. Throughout this paper, all appearing Matsubara frequencies are therefore positive.

The Bogoliubov--de Gennes symmetry of the hamiltonian leads to the self-conjugation constraint
\begin{equation}
Q=\tau_1\sigma_1Q^T\tau_1\sigma_1.\label{selfconjugate}
\end{equation}

To model a Josephson junction with a nonsuperconducting link of length $L$, we consider a steplike dependence of the superconducting order parameter on coordinate (see Fig.~\ref{fig1}):
\begin{equation}
	\check\Delta(x) =
	\begin{cases}
		\Delta \tau_{\phi}(x), & |x|>L/2, \\
		0, & |x| < L/2,
\end{cases}
\end{equation}
where
\begin{equation}
\tau_{\phi}(x) = \tau_1\cos{\hat\phi(x)}-\tau_2\sin{\hat\phi(x)},
\end{equation}
and $\hat\phi(x)$ is the superconducting phase. $\hat\phi(x)$ is constant within each superconducting lead, i.e., in a symmetric gauge $\hat\phi(x)=\sgn (x)\hat\varphi/2$ where $\hat\varphi$ is the phase difference of the Josephson junction.

The action Eq.\ \eqref{act1} describes the case of magnetic impurities. For simplicity we focus on this case in what follows. The other cases of ferromagnetism or external magnetic fields are treated very similarly, with minor alterations to the model and result. We will discuss these in detail in Sec.~\ref{secF}.

\section{Saddle-point configuration} \label{secSaddle}
To calculate the functional integral in Eq.\ \eqref{IQ}, we start with the saddle-point of the action Eq.\ \eqref{act1}, i.e., we find a configuration $Q_0(x)$ that extremizes $S[Q]$. The saddle-point solution has the form
\begin{equation}
Q_0(x) = \Lambda\cos\hat\theta(x) + \tau_{\phi}(x) \sin\hat\theta(x), \label{QLsaddle}
\end{equation}
where $\hat\theta(x)$ interpolates between the superconducting value
\begin{equation}
\hat\theta_0=\arctan \frac\Delta{\hat\omega}
\end{equation}
at $x\to\pm\infty$ and $\hat\theta\equiv0$ at $|x| < L/2$, where the superconducting components of $Q$ are suppressed by magnetism.

The variational equation on $\hat\theta(x)$ is identical to the Usadel equation. For a uniform superconductor with constant pair potential $\Delta$, it becomes a sine-Gordon equation:
\begin{gather}
\nabla^2\hat\theta - \hat\varkappa^2\sin(\hat\theta-\hat\theta_0) = 0, \label{UsadelEq}\\
\hat\varkappa^2 = \frac{2\sqrt{\hat\omega^2+\Delta^2}}{D}
\end{gather}
with the boundary conditions $\hat\theta(\pm\infty)=\hat\theta_0$ and $\hat\theta(\pm L/2)=0$. It is solved by
\begin{equation}
\hat\theta(x)=
\begin{cases}
\hat\theta_0 - 4\arctan\left[\tan\left(\frac{\hat\theta_0}{4}\right)e^{-\left(|x|-\frac{L}{2}\right)\hat\varkappa}\right],\ &|x|>\frac{L}2,\\
0, &|x|<\frac{L}2.
\end{cases}
\label{thetaEq}
\end{equation}
This solution is shown by the blue curve in Fig.~\ref{fig1}.

The correlator Eq.\ \eqref{IQ} can be calculated in the saddle-point approximation, where only $S[Q_0]$ is used to calculate the integral in Eq.\ \eqref{IQ}. However, this produces a trivial result. Indeed, $S[Q_0]$ does not depend on superconducting phases $\phi(x)$ in the leads. Consequently, $\langle I(\varphi_1)I(\varphi_2)\rangle=0$ and thus $I(\varphi)\equiv0$ in this approximation. This is expected -- the saddle-point approximation of the sigma model is controlled by the Usadel equation and therefore only captures physics on the level of averaged Green's functions. Thus, it only calculates the reducible part $\langle I(\varphi_1)\rangle \langle I(\varphi_2)\rangle$ of the current-current correlator. This reducible part is vanishingly small since $\langle I(\varphi)\rangle\propto e^{-L/l_*}$. Our sigma model does not resolve such exponential smallness and therefore we get zero current in the saddle-point approximation.

\begin{figure}[t]
\centering
\hspace*{-3pt}\includegraphics[width=0.485\textwidth]{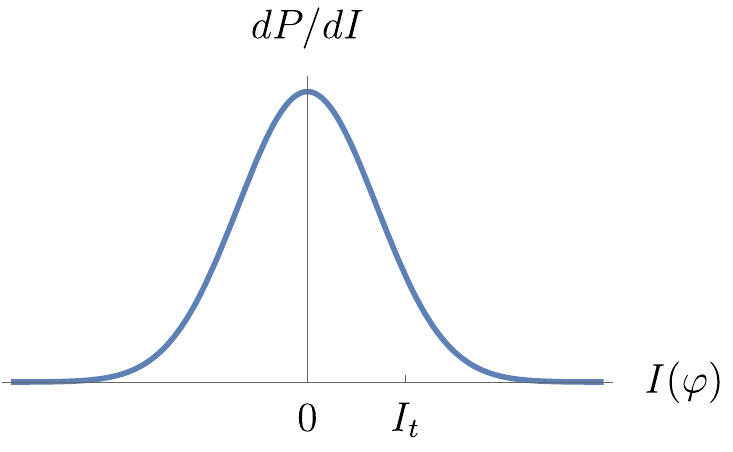}
\caption{
Probability distribution function of the sample-dependent supercurrent $I(\varphi)$. The current vanishes on average due to the random, sample-dependent sign, however its typical (r.m.s.) value $I_t$ is substantial.
}
\label{fig2}
\end{figure}

As will be shown below, the irreducible part of the correlator $\langle I(\varphi_1)I(\varphi_2)\rangle$ is not exponentially small. Thus, we have a situation where $\langle I(\varphi)\rangle\ll \sqrt{\langle I^2(\varphi)\rangle}$. This root mean square (r.\ m.\ s.) over disorder realizations turns out to be independent of $\varphi$.  The current is thus sample-dependent with a vanishing average value but with substantial fluctuations, as illustrated by Fig.~\ref{fig2}. In any particular sample, the measured supercurrent will typically be of the order of $I_t=\sqrt{\langle I^2\rangle}$.

\section{Fluctuations of $Q$ around the saddle-point}\label{secFluct}
To obtain a nonzero current-current correlator, we need to integrate the functional integral in Eq.\ \eqref{IQ} over fluctuations of $Q$ around $Q_0$. We parameterize these fluctuations by a small field $W(x)\ll1$ in the following way
\begin{equation}
Q(x) = e^{-\frac12 \hat\theta(x)\sigma_3\tau_3\tau_{\phi}(x)}\Lambda e^{iW(x)}e^{\frac12 \hat\theta(x)\sigma_3\tau_3\tau_{\phi}(x)}, \label{Wparam}
\end{equation}
with $\hat\theta(x)$ defined by Eqs.\ \eqref{thetaEq}. At $W=0$ the saddle-point configuration $Q_0(x)$ is restored. $W$ obeys
\begin{gather}
\{W,\Lambda\}=0,\label{Wsym1}\\
W= - \tau_1\sigma_1W^T\tau_1\sigma_1, \label{Wsym2}
\end{gather}
following from $Q^2=1$ and Eq.\ \eqref{selfconjugate}, respectively.

We now substitute Eq.\ \eqref{Wparam} into Eq.\ \eqref{act1} to obtain an action in terms of $W$. Linear terms vanish, since $Q_0$ extremizes $S[Q]$, and terms higher than quadratic in $W$ are neglected. Details of the calculation are found in Appendix~\ref{appS}, and the result is
\begin{multline}\label{actWs}
S_W^{(\mathrm{S})}= \frac{\pi\nu D}{8}\int dx \tr\biggl\{(\nabla W)^2 - \frac14[\tau_\phi\tau_3\sigma_3\nabla\hat\theta,W]^2
\\
+ \hat\varkappa^2\cos(\hat\theta_0-\hat\theta)W^2 \biggr\}
\end{multline}
in the superconducting leads, while in the magnetic part
\begin{equation}\label{actWm}
S_W^{(\mathrm{M})}= \frac{\pi\nu D}{8}\int dx \tr\biggl\{(\nabla W)^2
+ \frac{2\hat\omega}D W^2 \biggr\}.
\end{equation}

The constraints, Eqs.\ \eqref{Wsym1} and \eqref{Wsym2} resolve into the following structure of $W$ in Nambu and particle-hole spaces:
\begin{equation}
W = \begin{pmatrix}
d\sigma_1+d'\sigma_2 & 0 \\
0 & -d^T\sigma_1-d'^T\sigma_2
\end{pmatrix}. \label{WEq}
\end{equation}
Here the fields $d(x)$ and $d'(x)$ are unconstraint matrices in replica space. They correspond to diffuson degrees of freedom \cite{NoteDiffusons}, while off-diagonal components, representing cooperons, are suppressed at $|x|<L/2$ by magnetic impurities due to the constraint of Eq.\ (\ref{Qtau3}). {\color{black}Diffusons and cooperons are the only soft modes in a diffusive system. Diagrammatically, they are two-particle propagators made of two electron Green's functions travelling in the opposite (diffuson) or the same (cooperon) direction, connected by the disorder lines, i.e., a so-called ladder diagram, as seen in Fig.~\ref{FigLoop}.}

In the superconducting leads, the off-diagonal (cooperon) components in $W$ do exist, but they can neither mix with diffuson modes nor penetrate the magnetic region and are thus discarded. Plugging the matrix Eq.\ \eqref{WEq} into Eqs.\ \eqref{actWs} and \eqref{actWm}, and tracing out Nambu and particle-hole space, we get
\begin{gather}\label{SdS}
\begin{multlined}
S_d^{(\mathrm{S})} = \frac{\pi\nu D}{2} \int dx \tr \biggl\{
(\nabla d)^2  + \left(\hat\varkappa^2-(\nabla\hat\theta)^2\right)d^2
\\
+ \frac{1}{2}(\nabla\hat\theta) e^{-i\hat\phi}d(\nabla\hat\theta) e^{i\hat\phi}d^T\biggr\} + (d\rightarrow d') ,
\end{multlined}\\
S_d^{(\mathrm{M})} = \frac{\pi\nu D}{2} \int dx \tr \biggl\{
(\nabla d)^2  + \frac{2\hat\omega}D d^2\biggr\} + (d\rightarrow d') .
\end{gather}

\section{Calculation of $\left< I(\varphi_1)I(\varphi_2) \right>$}\label{secCorr}
The above action is a quadratic form: $S_d = d^{ij}(x)\mathcal{H}_{ij,kl}(x,x')d^{kl}(x') + (d\rightarrow d')$, and the gaussian integral in Eq.\ \eqref{IQ} is given by its determinant \cite{commentDet}:
\begin{equation}
Z=\int e^{-S[Q]}DQ=\int e^{-S_d[d,d']}DdDd'=\frac{1}{\det \mathcal{H}}. \label{Idet}
\end{equation}
To find this determinant we employ the Gelfand-Yaglom theorem \cite{GY}, which circumvents the calculation of individual eigenvalues and derives their product directly from the eigenvalue equation. If the eigenvalue equation is $\mathcal{F}(\lambda)=0$ then the following holds (for details and full requirements, see Ref.~\cite{Dunne}):
\begin{equation}
Z = \frac{\mathcal{F}(-\infty)}{\mathcal{F}(0)}.
\end{equation}
The eigenmode equation for a diffuson mode $d(x)$ involving replica indices $ij$ (assuming $i\neq j$) has the form of a Schr\"odinger equation on the spinor $(d_{ij}, d_{ji})^T$:
\begin{equation}
H_{ij}(x) \begin{pmatrix} d_{ij} \\ d_{ji}\end{pmatrix} = \frac{\lambda}{\pi\nu D} \begin{pmatrix} d_{ij} \\ d_{ji}
\end{pmatrix},\label{schr1}
\end{equation}
where
\begin{multline}
H_{ij}^{(\mathrm{S})}(x)=\\
\begin{pmatrix}  \frac{\varkappa_{i}^2 + \varkappa_{j}^2 - (\nabla\theta_i)^2 - (\nabla\theta_j)^2}2 - \nabla^2& \frac12(\nabla\theta_i)(\nabla\theta_j)e^{i\phi_i-i\phi_j}\\
\frac12(\nabla\theta_i)(\nabla\theta_j)e^{i\phi_j-i\phi_i} & \frac{\varkappa_{i}^2 + \varkappa_{j}^2 - (\nabla\theta_i)^2 - (\nabla\theta_j)^2}2-\nabla^2\end{pmatrix}\label{HS}
\end{multline}
in the superconducting leads and
\begin{equation}
H_{ij}^{(\mathrm{M})}(x)=(\omega_i + \omega_j)/D - \nabla^2\label{HM}
\end{equation} in the magnetic region.

To obtain $\mathcal{F}(\lambda)$ from the Schr\"odinger equation \eqref{schr1}, we construct the wave functions in three regions (magnetic region and two leads) and write down matching equations at $x=\pm L/2$. These form a linear set of equations, whose determinant satisfies the requirements set on $\mathcal{F}$. A detailed calculation is found in Appendix~\ref{appF} and yields
\begin{align}
\mathcal{F}_{ij}(0) &= 1 - f^2\cos^2\frac{\delta\varphi_{ij}}{2},\label{Fij}\\
\mathcal{F}_{ij}(-\infty) &= 1,\label{Fij}
\end{align}
where $\delta\varphi_{ij} = \varphi_{i} - \varphi_{j}$ is the difference between the superconducting phase differences $\varphi$ in replicas $i$ and $j$. The function $f$ in Eq.\ \eqref{Fij} is
\begin{equation}
f = \frac{q_+ - q_-}
{(q_++q_-)\cosh(kL) + \left(k+\frac{q_+q_-}{k}\right)\sinh(kL)}.\label{eqf}
\end{equation}
Here $k=\sqrt{(\omega_1+\omega_2)/D}$ and $q_\pm$ is the solution of
\begin{gather}
\begin{multlined}2\nabla q_\pm - 2q^2_\pm + \varkappa_i^2 + \varkappa_j^2 -\\
- (\nabla\theta_i)^2 - (\nabla\theta_j)^2 \mp(\nabla{\theta}_i)(\nabla{\theta}_j) = 0,
\end{multlined}\label{eqq}\\
q_\pm(x\to+\infty)>0,
\end{gather}
taken at the interface, $x=L/2$.
Equation \eqref{eqq} does not have a general analytical solution, but it can be solved in various limits and special cases. These are discussed in detail in the next Section.

Substituting the partition functions of the diffuson degrees of freedom into Eq.\ \eqref{IQ}, we get
\begin{multline}\label{corrIntermediate}
\langle I(\varphi_1)I(\varphi_2)\rangle
\\
= 4e^2T^2\sum\limits_{\omega_1,\omega_2}\frac{\partial^2 }{\partial\varphi_1\partial\varphi_2}\lim\limits_{n_{1,2}\to0} \frac{1}{n_1n_2}\prod\limits_{i,j}\frac{1}{\mathcal{F}_{ij}(0)}.
\end{multline}

Replica indices $i,j$ each run over two blocks of replica -- the first contains $n_1$ replica with $\omega=\omega_1,\ \varphi=\varphi_1$ and the second block has $n_2$ replica with $\omega=\omega_2,\ \varphi=\varphi_2$. Suppose $i,j$ are both from block $1$. The function $\mathcal{F}_{ij}(0)=\mathcal{F}_{11}(0)$ in this case does not depend on superconducting phase since $\delta\varphi_{11}\equiv0$. There are $n_1^2$ such $i,j$ pairs and in the replica limit $n_1\to0$ their contribution amounts to $\mathcal{F}_{ij}^{n_1^2}(0)\to1$. Thus, they are irrelevant and can be dropped from the product in Eq.\ \eqref{corrIntermediate}. As a result, only $F_{ij}$ with $i,j$ from different blocks are relevant and the replica limit $n_{1,2}\to0$ resolves into
\begin{multline}
\langle I(\varphi_1) I(\varphi_2) \rangle =
4e^2T^2\sum\limits_{\omega_1,\omega_2}\frac{\partial^2 \ln \mathcal{F}_{12}(0)}{\partial \delta\varphi_{12}^2}\\
=2e^2T^2\sum\limits_{\omega_1,\omega_2}f^2\frac{\cos\delta\varphi_{12}
-f^2\cos^2\frac{\delta\varphi_{12}}{2}}{(1-f^2\cos^2\frac{\delta\varphi_{12}}{2})^2}, \label{IF}
\end{multline}
The result Eq.\ \eqref{IF} is an even function of $\delta\varphi_{ij}$, and can be written as a series over cosines:
\begin{equation}
\langle I(\varphi_1)I(\varphi_2)\rangle = \sum\limits_{n=1}^\infty J_n^2\cos(n\delta\varphi_{12}). \label{CorrJ}
\end{equation}
While the coefficients $J_n$ can only be found analytically in certain limits, significant information on the current-phase relation statistics can be obtained without their explicit knowledge.

\begin{figure}[t]
\centering
\hspace*{-3pt}\includegraphics[width=0.485\textwidth]{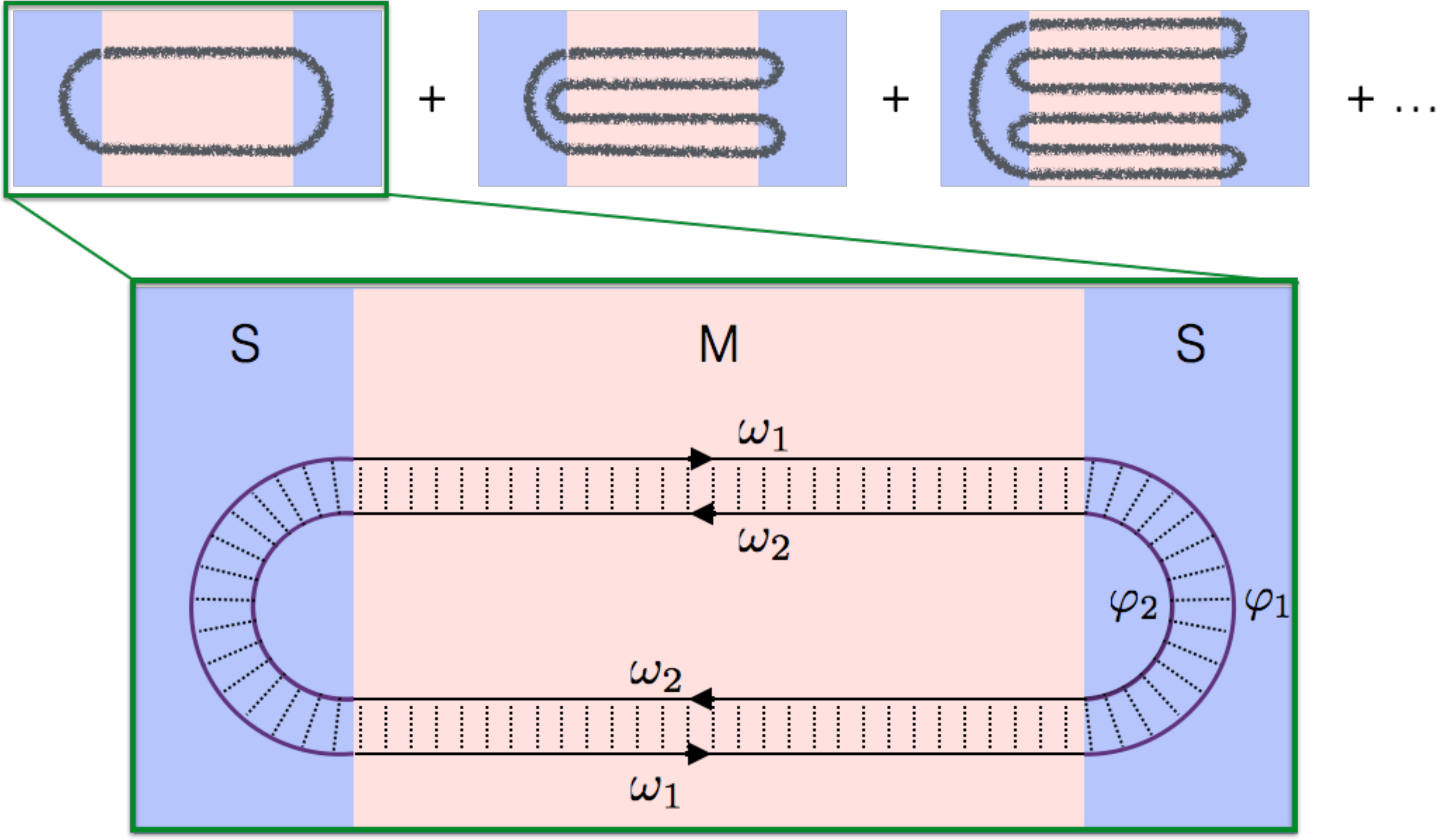}
\caption{
Feynman diagram for $\langle I(\varphi_1)I(\varphi_2)\rangle$. Rough lines represent diffusons. The first term is resolved in detail below the sum. The diffuson ladders in the S leads involve inhomogeneous matrix Green's functions incorporating the inverse proximity effect. The first diagram is proportional to $\cos\delta\varphi_{ij}$ and thus only contributes to the first harmonic. The second diagram has diffusons traversing M twice, allowing a $\cos(2\delta\varphi_{ij})$-term to emerge. This diagram contributes both to $J_1^2$ and to $J_2^2$ of Eq.\ \eqref{CorrJ}. The third diagram contributes to $J_1^2$, $J_2^2$, and $J_3^2$, etc.
}
\label{FigLoop}
\end{figure}

The above calculation is equivalent to taking a series of Feynman diagrams consisting of a single diffuson loop, see Fig.~\ref{FigLoop}. The diffuson can connect the S leads multiple times, which then contributes to higher harmonics. The Green's functions involved in the diagrams correspond to the saddle-point solution we found earlier, i.e., are governed by the Usadel equation. In particular, the Green's functions in S are inhomogeneous and have a nontrivial electron-hole structure. This structure allows them to connect two electron Green's functions that enter the lead with the same arrow direction (i.e., two electrons enter S), which is the diagrammatic representation of Andreev reflection.

\section{Current-phase statistics}\label{secStat}
The current-phase relation in the junctions we consider depends strongly on the particular disorder realisation and can be any $2\pi$-periodic function $I(\varphi)$. Let us expand it into harmonics via
\begin{equation}
I(\varphi) = \sum\limits_{n=1}^\infty \left[c_n\cos(n\varphi) + s_n\sin(n\varphi)\right]. \label{Ics}
\end{equation}
Here $c_n,s_n\in (-\infty,\infty)$ are sample-dependent coefficients representing the $n$-th harmonic component in $I(\varphi)$ of a particular junction. The statistics of the current-phase relations are thus contained in the statistics of $c_n$ and $s_n$. To study the latter, we substitute the series Eq.\ \eqref{Ics} into the left-hand side of Eq.\ \eqref{CorrJ}. We immediately find that
\begin{gather}
\langle c_nc_m\rangle = \langle s_ns_m\rangle = \delta_{nm}J_n^2,\\
\langle c_ns_m\rangle = 0.
\end{gather}
Furthermore, the probability distribution function of $\{c_n,s_n\}$ is gaussian. This does not follow directly from Eq.\ \eqref{Ics}, but rather from the action $S_W$ being gaussian. Indeed, higher-order current correlators are governed by the same gaussian action, except that the replica space is larger, containing as many replica blocks as there are currents in the correlator. Consequently, higher-order correlators reduce to second-order ones. Diagrammatically, this means that a higher-order correlator is simply a product of diffuson loops as in Fig.~\ref{FigLoop}, summed over possible permutations of replica indices among the loops. As a result $c_{1,2,\dots}$ and $s_{1,2,\dots}$ are independent gaussian random variables with zero mean and dispersion $J_n^2$:
\begin{equation}
dP(c_n,s_n) = \frac{1}{2\pi J_n^2} \exp\left( -\frac{c_n^2+s_n^2}{2J_n^2} \right) d c_n d s_n. \label{Pcs}
\end{equation}
Being a linear combination of $c_n$ and $s_n$, the current $I(\varphi)$ at any fixed phase difference $\varphi$ is also normally distributed with dispersion
\begin{equation}
I_t^2\equiv \sum_{n=1}^\infty J_n^2 = 2 e^2 T^2 \sum_{\omega_1,\omega_2} f^2.
\end{equation}

Each harmonic in Eq.\ \eqref{Ics} can also be parameterized as $a_n\sin(n\varphi - \varphi_{0n})$ in terms of sample-dependent amplitude $a_n\geqslant0$ and phase shift $\varphi_{0n}\in(0,2\pi]$. Their distribution follows from Eq.\ \eqref{Pcs}:
\begin{equation}
dP(a_n) = \frac{1}{J_n^2} \exp\left( - \frac{a_n^2}{2J_n^2} \right) a_n d a_n, \label{Paph}
\end{equation}
while $\varphi_{0n}$ is independently uniformly distributed. The latter fact means that the effective superconducting phase is completely randomized by the disorder. Breaking of TRS is essential here -- it eliminates the reference point for superconducting phase in the system. This destruction of superconducting phase memory is of the same nature as the suppression of the average current $\langle I(\varphi)\rangle$. Indeed, the average current is proportional to $\sin(\varphi)$ without any phase shift. It is only the fluctuational part that we study here that exhibits the random phase shift.
Note that the breaking of TRS does not need to contain randomness itself: a uniform magnetic field in conjunction with potential impurity scattering achieves the very same effect.

An important simple case is when $J_{n>1}=0$, so that only the first harmonic remains. As we will show, this happens at relatively high temperatures, when $T$ exceeds either $\Delta$ or $E_\mathrm{Th}$. In this case,
\begin{equation}
I(\varphi) = I_c\sin(\varphi-\varphi_0),
\end{equation}
with uniformly distributed phase shift and a critical current $I_c=a_1$ being distributed according to Eq.\ \eqref{Paph}, with an average $\langle I_c\rangle = J_1\sqrt{\pi/2}$.

When multiple harmonics are present, the distribution function of $I_c\equiv \max_\varphi I(\varphi)$ becomes complicated. Instead, we use $I_t$ to characterize the typical magnitude of the Josephson effect.

\begin{figure}[t]
\centering
\hspace*{-3pt}\includegraphics[width=0.485\textwidth]{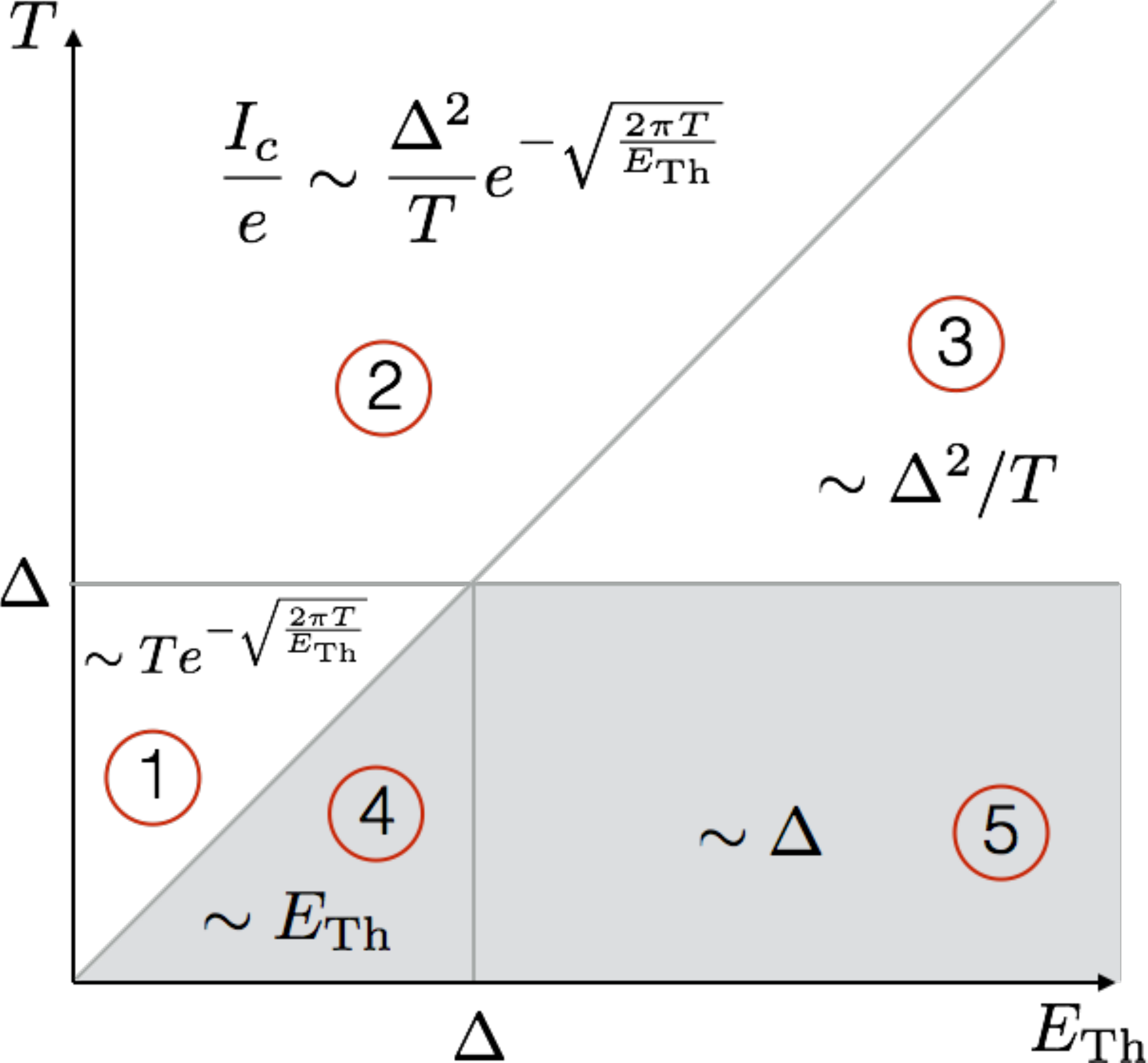}
\caption{
Diagram of the Josephson current in different cases. Numbers refer to the limits listed in Table~I. Higher harmonics of the supercurrent $I(\varphi)$ are present in the shaded region where temperature is the smallest energy scale.
}
\label{FigDiagram}
\end{figure}

We will now calculate $J_n$ and $I_t$ in different physical limits shown in the diagram, Fig.~\ref{FigDiagram}, collecting the results in Table~I. The table lists cases, values of $I_t$ and the harmonics content of $I(\varphi)$ in each case. Details of the calculations are available in Appendix~\ref{appII}.

\begin{table}[h]
\caption{Fluctuational current results} % title of Table
\centering % used for centering table
\begin{tabular}{c |c | c| c} % centered columns (4 columns)
\hline\hline %inserts double horizontal lines
& Limit & $I_t$ & harmonics\\ [0.5ex] % inserts table
%heading
\hline % inserts single horizontal line
1. & $\Delta\gg T\gg E_\mathrm{Th}$ & $\phantom{\Biggl|}\frac{2\sqrt{2}eT}{\pi^2}\exp\left[-\sqrt{\frac{2\pi T}{E_\mathrm{Th}}}\right]\phantom{\Biggl|}$ & $J_n=\delta_{1n}I_t$ \\ % inserting body of the table
\hline
2. & $T\gg E_\mathrm{Th};\Delta$ & $\frac{e\Delta^2}{4\sqrt{2}\pi^2 T}\exp\left[-\sqrt{\frac{2\pi T}{E_\mathrm{Th}}}\right]\phantom{\Biggl|}$ & $J_n=\delta_{1n}I_t$\\ % inserting body of the table
\hline
3. & $E_\mathrm{Th}\gg T\gg \Delta$ & $\phantom{\Biggl|}1.15\cdot\frac{e\Delta^2}{4\sqrt{2}\pi^2 T}$ & $J_n=\delta_{1n}I_t$ \\ % inserting body of the table
\hhline{=|=|=|=}
4. & $\Delta\gg E_\mathrm{Th}\gg T$ & $\phantom{\Biggl|}\sqrt{\frac{3}{2}\zeta(3)}\frac{eE_\mathrm{Th}}{\pi}\phantom{\Biggl|}$ & $J_n = \frac{I_t}{n^{3/2}\sqrt{\zeta\left(\frac32\right)}}$ \\ % inserting body of the table
\hline
5. &\begin{tabular}{@{}c@{}} $E_\mathrm{Th}\gg \Delta\gg T$  \\ $(L\to0, T\to0)$\end{tabular}
& $\phantom{\Biggl|}0.101e\Delta\phantom{\Biggl|}$ &
\begin{tabular}{@{}c@{}}$J_1=0.908I_t$  \\ $J_2=0.325I_t$ \\ $J_3=0.181I_t$ \\ $J_4=0.119I_t$ \\ $\vdots$\end{tabular}\\ % inserting body of the table
\hline %inserts single line
\end{tabular}
\end{table}

\subsection{High temperature}
When temperature is larger than the Thouless energy, $T\gg E_\mathrm{Th}$, diffusons decay over a thermal length that is shorter than the junction length. This produces an exponentially small $f\propto \exp[-\sqrt{(\omega_1+\omega_2)/E_\mathrm{Th}}]$. For $f\ll1$, the summand in Eq.\ \eqref{IF} simplifies to $f^2\cos\delta\varphi_{12}$, meaning that only the first harmonic survives. Furthermore, the lowest term in the Matsubara sum, with $\omega_1=\omega_2=\pi T$ dominates the sum due to the exponential factor. At equal frequencies, $\omega_1=\omega_2$, Eq.\ \eqref{eqq} is solved exactly by $q_-=-\nabla^2\theta/\nabla\theta$ and $q_+=-\nabla^3\theta/\nabla^2\theta$, and we find
\begin{equation}\label{eqIc1}
I_t = 2\sqrt{2}eT\left(\frac{\sqrt{1+\sqrt{1+\left(\frac{\Delta}{\pi T}\right)^2}}-\sqrt{2}}{\sqrt{1+\sqrt{1+\left(\frac{\Delta}{\pi T}\right)^2}}+\sqrt{2}}\right) e^{-\sqrt{\frac{2\pi T}{E_\mathrm{Th}}}}.
\end{equation}
The current-phase relation is sinusoidal,
\begin{equation}
I(\varphi) = I_c\sin(\varphi-\varphi_0),
\end{equation}
with random, uniformly distributed phase shift $\varphi_0$ and a critical current with $\langle I_c\rangle = I_t\sqrt{\pi/2}$, distributed according to Eq.\ \eqref{Paph} (with $I_c=a_1$).

The result Eq.\ \eqref{eqIc1} covers the upper left triangle of the diagram and can further be divided into the sublimits~1 where $\Delta\gg T$ and~2 where $\Delta\ll T$. The corresponding expressions for $I_t$ are listed in Table~I.

Another high temperature case is $E_\mathrm{Th}\gg T\gg \Delta$, case~3 on the diagram. This means a short junction and a temperature high only with respect to $\Delta$. In this case the small parameter is $\Delta/\omega_{1,2}$. Again $f\ll1$, this time because $q_+\approx q_-\approx k$ and $q_+-q_-\ll k$. Thus only the first harmonic survives. Eq.\ \eqref{eqq} can be solved perturbatively, and the sum over $\omega_{1,2}$ is taken. The result for $I_t$ is written in Table~I.

\subsection{Low temperature}
At low temperatures, $T\ll \Delta, E_\mathrm{Th}$, we replace Matsubara sums in Eq.\ \eqref{IF} by integrals (which corresponds to putting $T=0$). The only parameter left in this case is $\Delta/E_\mathrm{Th}$.

Let us first consider the long junction limit, $\Delta \gg E_\mathrm{Th}$, which is case 4 of the diagram. The integral over frequencies is effectively cut off at $\omega_{1,2}\sim E_\mathrm{Th}$, since higher-frequency terms are exponentially small. This means that relevant frequencies are all much smaller than $\Delta$. Taking advantage of the small parameter $\omega_{1,2}/\Delta$, we can then solve Eq.\ \eqref{eqq} perturbatively and take the Matsubara integral, yielding:
\begin{equation}\label{Jn}
J_n^2 = \frac{3e^2E^2_\mathrm{Th}}{2\pi^2n^3}.
\end{equation}
All harmonics are present in this case, though their magnitude decays quickly with their number $n$: $J_n\propto n^{-3/2}$. The current scales as $eE_\mathrm{Th}$, similarly to the average Josephson current in a conventional long SNS junction. The expression for the typical current is written in the table, case 4.

\begin{figure}[t]
\centering
\hspace*{-3pt}\includegraphics[width=0.485\textwidth]{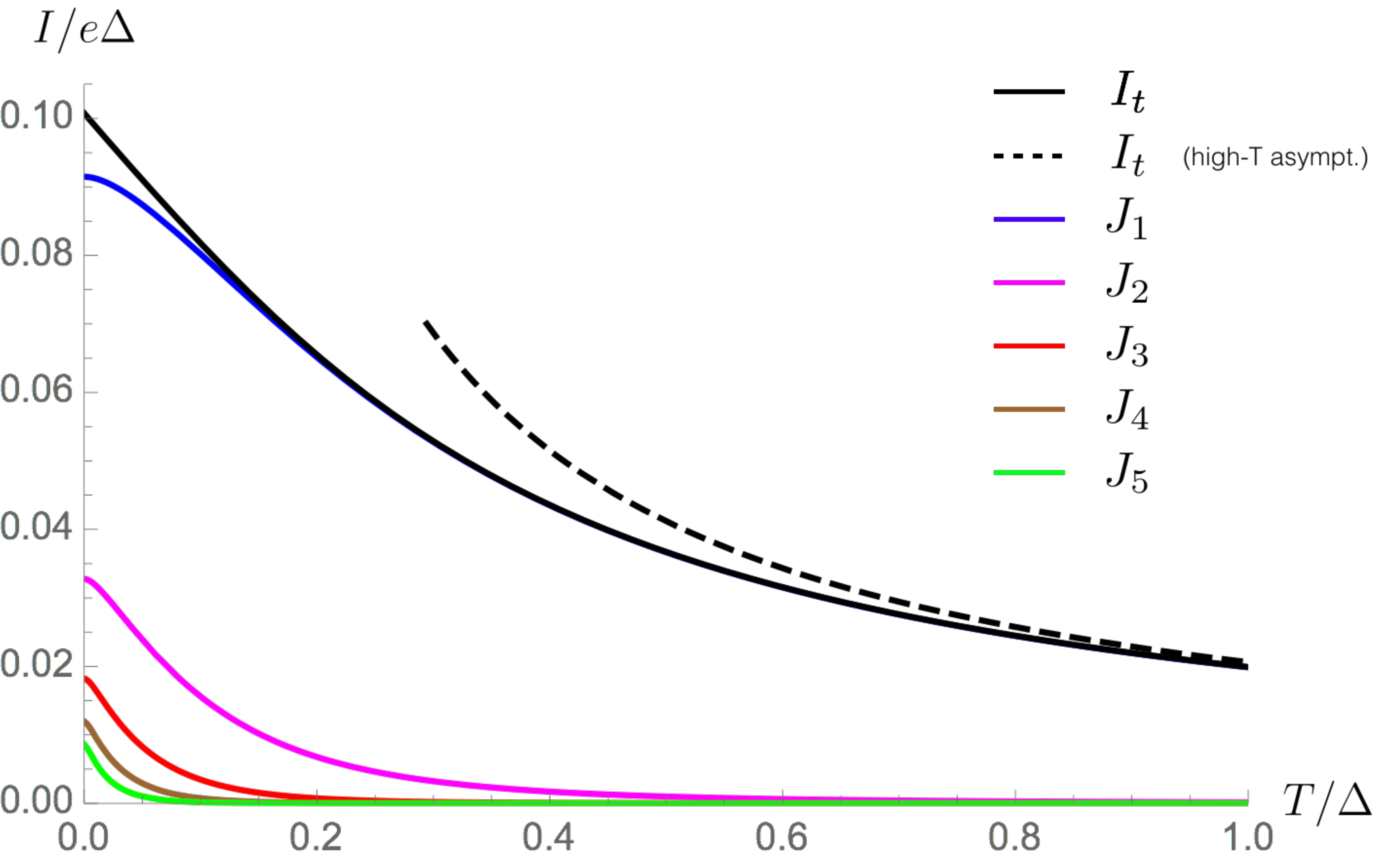}
\caption{
Current in a short junction. The solid black curve shows the typical current $I_t$. The curves below are $J_n$ with harmonic number increasing from top to bottom. Higher harmonics, $J_{n>1}$, fall off quickly with temperature, so that $I_t$ becomes identical to the first harmonic $J_1$. The dashed hyperbole is the high-temperature asymptotic for $I_t$, case 3 in Table~I.
}
\label{figShort}
\end{figure}

In the opposite limit of a short junction, $E_\mathrm{Th}\gg\Delta\gg T$, the Matsubara integral converges over $\omega_{1,2}\sim\Delta$ meaning that $kL\ll1$ for all relevant frequencies and can be neglected. Thus there are no parameters remaining, and the only energy scale is $\Delta$, so that inevitably $I\sim e\Delta$. Nevertheless, Eq.\ \eqref{eqq} cannot be solved analytically in this case (it requires solving for all $\omega_{1,2}/\Delta$). We treated this case numerically, with the result for $T\to0$, $E_\mathrm{Th}\to\infty$ presented in the table, case~5. Remarkably, while $I_t$ and individual harmonics $J_n$ cannot be found analytically, the derivative $\partial I_t^2/\partial T$ can be calculated, because it converges at small frequencies, where perturbation theory can be used. It yields
\begin{equation}
\left.\frac{\partial I_t^2}{\partial T}\right|_{T,L=0} = -\frac{e^2\Delta^2}{4\pi}.
\end{equation}
Peculiarly, while $I_t^2$ is linear in $T$, and $I_t^2 = \sum J_n^2$, each of the individual terms $J_n^2$ is quadratic in $T$. This is illustrated by Fig.~\ref{figShort} showing $I_t$ and $J_{1,2,3,4,5}$ obtained numerically.

\section{Ferromagnetic case}\label{secF}
In the derivation above we explicitly considered the case of a metallic link with magnetic impurities (case M). In this section we will discuss other magnetic links: SFS junctions (case F) as well as SNS junctions subject to an external magnetic field (case A).

Let us first discuss how different magnetic phenomena affect diffusive transport in the link. In the absence of any magnetic effects, electrons and holes propagate in the same way, i.e., electron and holes share the same diffusive trajectories due to TRS. In addition, these trajectories do not depend on the spin due to full spin symmetry. This gives rise to eight soft modes -- four diffusons and four cooperons, corresponding to the four possible spin configurations of the involved pair of fermions.

When magnetic impurities (i.e., a random exchange field) are added, both TRS and spin symmetry are broken. The former means that all cooperons are suppressed. The latter means that three of the four diffusons are suppressed, with only the singlet diffuson surviving. The Feynman diagrams of Fig.~\ref{FigLoop} are made of this diffuson and constitute the $\langle I(\varphi_1)I(\varphi_2)\rangle$ correlator we calculated in the previous sections for the M case.

Next, consider the orbital effect of magnetic fields: a nontrivial vector potential $\mathbf{A}$ leads to different diffusion of electrons and holes. This breaks TRS and suppresses all four cooperons. The diffusons are unaffected by $\mathbf{A}$. Thus, if only orbital effects are present, the current-current correlator is four times that of the M case, and the typical current is two times larger: $I_A=2I_M$. This simple relation can also be deduced directly from the observation that the system in the A case consists of two identical subsystems (sharing the same disorder realization) with different spin. Each of these subsystems is equivalent to the M case, therefore the current is simply doubled.

Finally, there is the case F of a uniform exchange field. The hamiltonian $H=H_0 + hs_z$ obeys
\begin{equation}
H=H^*\label{TRS}
\end{equation}
\cite{NoteTRS}. This is a spinless TRS and indicates the survival of some cooperon modes. Indeed, the uniform exchange field produces a Zeeman energy shift between the spin-up and spin-down subsystems. The diffusion in those subsystems becomes effectively uncorrelated due to this shift, meaning that soft modes pairing particles with different spin are suppressed, while same-spin modes are unaffected.
{\color{black} Thus, two diffusons and two cooperons are present in this case \cite{NoteFAverage}.}

In addition to the exchange field, a ferromagnet always has a magnetic field inside producing an orbital effect. The two effects act on different length scales: the exchange field {\color{black} breaks pairs on the} length $l_h=\sqrt{D/h}$, while the orbital effect involves a larger scale $l_A$. Thus, there are two cases -- the junction can be either shorter or longer than $l_A$.

In the first case, F-long, the junction is long in the sense $L\gg l_A,l_h$, so that both suppression mechanisms are relevant, only leaving two diffusons intact. Thus, the correlator it twice that of the M case:
\begin{equation}
\langle I_{Fl}(\varphi_1)I_{Fl}(\varphi_2)\rangle = \sum\limits_{n=1}^\infty 2J_n^2\cos(n\delta\varphi_{12}) \label{CorrJFLong}
\end{equation}
or simply $I_{Fl}=\sqrt{2} I_M$. While in the M case we had two \textit{identical} subsystems with different spin, in the F-long case we have two \textit{uncorrelated} subsystems with different spin. In the latter case we thus have to double the current-current correlator instead of the current, producing the above $\sqrt{2}$ factor.

Finally, in the case F-short, the junction is of intermediate length, $l_A\gg L\gg l_h$. In this case exchange effects are strong enough to suppress the average current, but orbital effects are irrelevant. In this case a cooperon contribution must be added to the diffuson contribution Eq.\ \eqref{CorrJFLong}. It can be found by enabling off-diagonal terms in the fluctuation matrix $W$, Eq.\ \eqref{WEq} and calculating the partition function of these new degrees of freedom. Adding the cooperon contribution to the diffuson contribution, we obtain
\begin{equation}
\langle I_{Fs}(\varphi_1)I_{Fs}(\varphi_2)\rangle = \sum\limits_{n=1}^\infty 4J_n^2\sin n\varphi_1\sin n\varphi_2. \label{CorrJFerroShort}
\end{equation}
This result can be obtained directly from symmetry consideration without explicit sigma-model calculations. Indeed, the cooperon contribution is a function of $\varphi_1+\varphi_2$ while the total $\langle I_{Fs}(\varphi_1)I_{Fs}(\varphi_2)\rangle$ can only contain $\sin m\varphi_1\sin n\varphi_2$ terms to satisfy $I(\varphi)=-I(-\varphi)$ required by Eq.\ \eqref{TRS}. These requirements fully fix the cooperon contribution and result in Eq.\ \eqref{CorrJFerroShort}.

Note that the cases M, A, F-long are very similar -- the only difference is an overall factor in the current magnitude, i.e.,
\begin{equation}\label{currentprop}
I_A=\sqrt{2}I_{Fl}=2I_M.
\end{equation}
The F-short case is different due to the spinless TRS Eq.\ \eqref{TRS}, which dictates an odd current-phase relation. This obviously means different statistics. Indeed, we have
\begin{equation} \label{IFs}
I_{Fs}=\sum_n s_n\sin(n\varphi),
\end{equation}
where $s_n$ are normally distributed with
\begin{equation} \label{snsm}
\langle s_n s_m\rangle = 4J_n^2\delta_{nm},
\end{equation}
where $J_n$ are still given by the same Table~I, as in the M case. There are no arbitrary phase shifts in $I_{Fs}$, 
{\color{black} only random signs of amplitudes $s_n$.} 
Interestingly, this does not turn the F-short junction into a $0$ or $\pi$ junction. The TRS does require that $\varphi=0,\pi$ be local extrema of the energy, however, 
{\color{black}
the global energy minima might lie elsewhere (but symmetrically with respect to 0, i.e., at $\pm \varphi$ with $\varphi \neq 0,\pi$).
This represents the situation known as $\varphi$ junction \cite{Buzdin2003,Goldobin2007}. 
At the same time, while known realizations of the $\varphi$-junction state are based on structures with alternating $0$ and $\pi$ junctions \cite{Buzdin2003,Goldobin2007}, our results demonstrate that fluctuational regime opens up a simpler way to implement it. While the average current in the F-short case can only correspond to either $0$ or $\pi$ junction, a specific realization of the system can be in the $\varphi$-junction state due to random signs of the Josephson harmonics.
}

Above we introduced the length scale $l_A$ of the orbital effects. It can be expressed through the magnetic field $\mathbf{B}$ and geometric parameters of the link \cite{we2017}. If the transverse sizes $w_y, w_z\ll L$ of the quasi-one-dimensional link are of the same order $w_y\sim w_z\sim w$, then $l_A = \Phi_0/(Bw)$ where $\Phi_0$ is the flux quantum. If the link has a strip geometry, i.e., $w_y\gg w_z$ then $l_A^{-2} = (B_z^2w_y^2+B^2_\parallel w^2_z)/\Phi_0^2$. Naturally, the in-plane magnetic field $B_\parallel$ has a weaker decoupling effect on the cooperons.

\section{Discussion}\label{secDisc}
From the diagram, Fig.~\ref{FigDiagram}, and Table~I we see that the Josephson current grows {\color{black}as temperature and junction length decrease, so that maximal current is achieved in the short-junction low-temperature limit, corresponding to the bottom right of the diagram. In this limit,the typical Josephson current $I_t$ is of the order of $e\Delta$. At temperatures higher than $E_\mathrm{Th}$ or $\Delta$, a small factor appears in the current, $e^{-\sqrt{T/E_\mathrm{Th}}}$ or $\Delta/T$, respectively, suppressing the supercurrent. Amplitudes $J_{n}$ of $n$-th harmonics contain $n$-th power of this small parameter. Therefore $I(\varphi)$ at high temperatures is dominated by the first harmonic and is therefore sinusoidal}. At low temperatures, in contrast, higher harmonics $J_n$ do not contain a small parameter. This means that in the low-temperature regime (shaded region on the diagram), the function $I(\varphi)$ can in principle be of arbitrary shape (it must only obey $\int_0^{2\pi} I(\varphi)d\varphi=0$). At the same time, $J_n$ decreases with $n$, so the main contribution to the current typically comes from the first harmonic. However, samples where the amplitude $a_1$ of the first harmonic is relatively small are not rare. For example, the probability for a sample to have $a_1<a_2$, i.e., for the second harmonic to exceed the first one, can be calculated from Eq.\ \eqref{Paph} and equals $P(a_1<a_2)=J_2^2/(J_1^2+J_2^2)$, which is $\approx11\%$ in the low-temperature limit (cases 4 and 5 in Table~I).

Thus, by creating multiple samples one may obtain one with a weak first harmonic. However, a simpler route is to work with a single sample and reshuffle the disorder until a configuration with a small $a_1$ appears. Such reshuffling can be achieved e.g. by changing the chemical potential via a back gate, or changing the strength or direction of an external magnetic field.

{\color{black} Josephson junctions with higher harmonics in the current-phase relations are long-known both theoretically and experimentally \cite{GolubovReview2004,DellaRocca2007,Robinson2007,Stoutimore2018}.} What distinguishes the system we study is that the relations between the different harmonics are random, sample-dependent. Not only are the amplitudes $a_n$ mutually uncorrelated, but also the phases are independent, allowing arbitrary current-phase relations.

The phase shifts in the fluctuational current are its most important feature. In all the cases we studied, 
{\color{black} except F-short [see discussion of this case below Eqs.\ (\ref{IFs}) and (\ref{snsm})],}
the current $I(\varphi)$ contains random phase shifts, making the system a $\varphi_0$ junction, i.e., a junction where current at $\varphi=0$ is nonzero and the energy minimum of the junction is at some $\varphi_0\neq0$ instead.

{\color{black} Previous research on $\varphi_0$ junctions has mainly focused on}
systems where the Josephson current is dominated by its disorder average $I(\varphi)=\langle I(\varphi)\rangle$ and fluctuations are irrelevant. For the average supercurrent to exhibit a $\varphi_0$ phase shift, the system must break the {\color{black} TRS, including its spinless version, Eq.\ \eqref{TRS}}, on average (i.e., not only for individual samples, but also for the disorder-averaged hamiltonian $\langle H\rangle$). Therefore, prospective $\varphi_0$-junction designs involve some ingredients required to fully break TRS. {\color{black} For example, a uniform exchange field is not sufficient to break the spinless TRS \cite{NoteTRS}. Full TRS breaking
in SFS junctions can be achieved via spin-orbit interaction \cite{Buzdin2008,Bergeret2015,Mironov2015a} or noncoplanar magnetization distribution in the ferromagnetic part \cite{Braude2007,Grein2009,Liu2010,Margaris2010,Kulagina2014,Moor2015,Silaev2017}. Another approach is to use quantum dots with spin-orbit interaction and applied exchange field \cite{Zazunov2009}, which has recently been implemented experimentally \cite{Kouwenhoven2016}. The necessary symmetry breaking due to spin-orbit interactions and exchange fields is also predicted to produce the $\varphi_0$-junction state in setups involving quantum point contacts \cite{Reynoso2008}, topological insulators \cite{Tanaka2009,Dolcini2015,Bobkova2016}, and nanowires \cite{Yokoyama2014,Mironov2015,Campagnano2015,Nesterov2016}.}

{\color{black} The fluctuational regime provides a different way of producing a $\varphi_0$ junction. One can use a simpler setup, e.g., a single-domain SFS (without spin-orbit interaction), where symmetry of the disorder- averaged system dictates $0$ or $\pi$ junction behavior for the average current. If such a system is in the fluctuational regime F-long (i.e., the junction is long compared to both exchange and orbital pair-breaking lengths $l_h$ and $l_A$), then the supercurrent is defined by the particular disorder realization, which breaks all symmetries (spinless TRS in particular), leading to $\varphi_0$-junction behavior. The advantage of this setup is its simplicity. The drawback is that the fluctuational current is small and does not scale with system size. It is also unpredictable by nature, only following the probability distribution we established.
}

Our results were obtained in the limit of q1D junctions, that is, for junctions of small-area cross section. Technically, this means that in general expressions for the diffusive modes in the wire \cite{Altshuler1987}, one should keep only zero transverse wave vector $q_\perp$.
Nonzero values of $q_\perp$ are of the order of $1/w$, the inverse width of the junction, and become important if $D q_\perp^2 \lesssim \omega_*$. Here, $\omega_* \sim \max \left( T, \min(\Delta,E_\mathrm{Th}) \right)$ is the characteristic Matsubara frequency at which the sum in Eq.\ (\ref{IF}) converges. The condition of quasi-one-dimensionality is then
\begin{equation}
w \ll \min \left( \sqrt{\frac DT}, \max \left( \sqrt{ \frac D\Delta},L \right) \right).
\end{equation}

SFS junctions with strong superconductors (large $\Delta$) have been previously studied in Ref.\ \cite{Altshuler1987}. They correspond to cases 1 and 4 of our diagram Fig.~\ref{FigDiagram}. We do agree with the general approach of Ref.\ \cite{Altshuler1987} and the qualitative dependences of our results on parameters coincide. However, there are certain discrepancies in numerical coefficients and subleading factors. We believe this is due to some arithmetical mistake in the central Eq.\ (20) of Ref.\ \cite{Altshuler1987}. 
{\color{black} We should also note certain inconsistency in Ref.\ \cite{Altshuler1987}, where Eq.\ (20) does not actually produce Eqs.\ (3) and (4).}

{\color{black}In the present paper, we assumed fully transparent interfaces between the magnetic link and the superconducting leads. This might be feasible for a single-nanowire setup where superconducting parts are simply proximitized segments of the nanowire, as in one-dimensional topological superconductor setups \cite{LutchynReview2018}. However, it is still important to understand whether high interface transparency $T_i$ is crucial and what happens for bad interfaces. We expect the supercurrent to decrease as transparency is decreased, and believe our results hold qualitatively as long as transparency is not much smaller than unity (e.g., $T_i \simeq 1/2$).
{\color{black} Higher harmonics should fall off quicker with decreasing the transparency.}
These conjectures are supported by results for the tunnelling limit, Ref.\ \cite{we2017}, where the current scales as $I\propto T_i^2$ {\color{black} at $T_i\ll 1$.}
%with $G_t$ being the tunnelling conductance of a single interface.
Higher harmonics in that limit contain higher powers of $T_i$ and are therefore negligible.
}

\section{Conclusions}\label{secConclusion}
We have studied the supercurrent $I(\varphi)$ occurring in SNS junctions where the average current is suppressed by magnetism in the N region. In such systems the supercurrent occurs due to mesoscopic fluctuations and is therefore highly sample-dependent, with random amplitude and phase shift. The distribution function of these random parameters has been found for various relations between $T,\Delta$ and Thouless energy $E_\mathrm{Th}$.

The fluctuational current does not scale with the number of conducting channels in the quasi-one-dimensional junction (similarly to universal conductance fluctuations) and is highest at low temperatures, $T\ll \Delta, E_\mathrm{Th}$. In this case it is typically of the order of $e\Delta$ or $eE_\mathrm{Th}$, whichever is smaller; the current-phase relation $I(\varphi)$ contains all harmonics. Amplitudes and phase shifts of individual harmonics are distributed independently;
{\color{black} the system is therefore in the $\varphi_0$-junction state (demonstrating anomalous Josephson effect with nonzero current at $\varphi=0$ and energy minimum shifted to $\varphi_0$).}
Typically, higher harmonics content is small ($<10\%$ of the total current). However, systems where the first harmonic is weaker than higher ones are not rare. At high temperatures, when either $T\gg E_\mathrm{Th}$ or $T\gg \Delta$, the current is sinusoidal (with random phase shift) and small in magnitude. These results are captured by Table~I and the diagram, Fig.~\ref{FigDiagram}.

All results are the same up to factors of $\sqrt{2}$ for ferromagnetic links, links with magnetic impurities, or links subject to strong magnetic fields. Relatively short SFS contacts ($l_h\ll L\ll l_A$) present a special case -- due to spinless time-reversal symmetry the random phase shifts in the harmonics of $I(\varphi)$ are restricted to random signs.
{\color{black} The system then can realize a $\varphi$ junction with two symmetric energy minima at phases $\pm\varphi$.}

From an experimental point of view, the limit $T\ll\Delta\ll E_\mathrm{Th}$ is the most interesting and relevant. This is the limit of a short junction at low temperature (the junction length should still be much larger than the {\color{black} pair-breaking} length $l_*$, so that the fluctuational current dominates). In this case the typical current magnitude is $I_t = 0.1e\Delta$ which is large enough to be observable.

\acknowledgments
We thank M.~V.\ Feigel'man for useful discussions.
This work was supported by the Russian Science Foundation (Grant No.\ 16-42-01035).
{\color{black} PAI was supported by the Basic research program of HSE and the 2018 RAS program.}

%==================APPENDIX====================

\appendix

\section{Full sigma model} \label{appModel}
The full action, including all possible magnetic terms, is
\begin{multline}\label{appSmain}
S = \frac{\pi\nu}{8} \int dx \tr \left\{
D (\nabla Q)^2 -\frac{D}{l_s^2} [\tau_3 \mathbf s,  Q]^2 \right.\\\left. -\frac{D}{l_A^2} [\tau_3, Q]^2
- 4 \left(\hat\omega\Lambda + \check\Delta + i\mathbf{hs}\tau_3\right)  Q\right\} .
\end{multline}
In addition to Nambu $(\tau_i)$ and particle-hole $(\sigma_i)$ spaces, Eq.\ \eqref{appSmain} also involves spin space $(s_i)$. The self-conjugation constraint on $Q$ is
\begin{equation}\label{appSelfconjugate}
Q = \tau_1 \sigma_1 s_2 Q^T \tau_1 \sigma_1 s_2.
\end{equation}

The $l_s$-term in action Eq.\ \eqref{appSmain} comes from the exchange interaction of electrons with magnetic impurities, averaged over positions and orientations of the latter. It leads to suppression of off-diagonal Nambu components in $Q$, as well as suppression of any nontrivial spin structure. Both suppressions happen on a length scale of $l_s$, meaning that $[Q,\tau_3]=0$ and $[Q,\mathbf{s}]=0$ in most of the magnetic region assuming $l_s\ll L$. The former is implemented in the main text by imposing Eq.\ \eqref{Qtau3} in the whole magnetic region. Spin suppression can be implemented in the same way, demanding that $Q\propto s_0$ in the whole magnetic region. However, $Q$ could still have nontrivial spin components in S, where magnetic impurities are absent. As we will show below, these components indeed show up in fluctuations of $Q$ around the saddle-point solution $Q_0$ (which itself commutes with $\mathbf{s}$). However, these fluctuational modes can neither penetrate the magnetic region, nor mix with spin-trivial modes and therefore are irrelevant to the Josephson effect. These considerations allow to simplify the model by setting $[Q,\mathbf{s}]=0$ in the whole system. After tracing spin out, we then restore the action Eq.\ \eqref{act1} and self-conjugation constraint Eq.\ \eqref{selfconjugate}. Note, however, that $\nu$ in Eq.\ \eqref{appSmain} is the metallic density of states per spin projection (since spin is explicitly resolved by the model) while $\nu$ in Eq.\ \eqref{act1} represents the total density of states since magnetic disorder mixes spin projections.

Note that it does not matter whether other magnetic terms ($h$-term and $l_A$-term) are also present at the same time since they equal zero when $[Q,\tau_3]=0,[Q,\mathbf{s}]=0$. In terms of symmetries, the $l_s$-term already breaks both spin symmetry and TRS, driving the sigma model in M to class A, which has no symmetries. Thus, adding further terms does not affect the sigma model any more.

The $l_A$-term in Eq.\ \eqref{appSmain} describes orbital effects produced by the vector potential $\mathbf{A}$. Initially, $\mathbf{A}$ enters the sigma model through the gradient term, where $\nabla$ is replaced by the long derivative $\nabla - i\mathbf{A}e/c[\Lambda,\cdot]$. The brackets here denote commutation with $\Lambda$. This leads to a linear term $S_{A1}\sim A$ and a quadratic term $S_{A2}\sim A^2$ in the action. Assuming a uniform magnetic field $\mathbf{B}$ for simplicity, we employ the Landau gauge with $\mathbf{A}\propto By$, where $y$ is the transverse coordinate of the link. Since we consider quasi one-dimensional wires, $Q(x,y)=Q(x)$. Thus, $S_{A1}$ vanishes after averaging over $y$. Only the quadratic term remains, $S_{A2}\propto \nu D\langle A^2\rangle_y\int\tr [\tau_3,Q]^2dx$ so that the decay length is $l_A\sim\Phi_0/(Bw)$ where $\Phi_0$ is the flux quantum and $w$ is the width of the link. A random vector-potential generates the same type of term (in this case the term linear in $\mathbf{A}$ vanishes when averaged over the ensemble).

The $\mathbf{hs}$-term corresponds to an exchange field $\mathbf{h}$. If this field is uniform, it reduces spin rotation symmetry to just $s_z$-symmetry. If it varies in direction, it breaks spin symmetry completely. However, as long as exchange fields in the link are coplanar, e.g. $\mathbf{h}(x)=(h_x(x),0,h_z(x))$ they do not fully break TRS. Indeed, the spinless TRS $H=H^*$ is preserved in this case. However, along with an exchange field $\mathbf{h}$, any ferromagnet also produces vector potentials. The corresponding $l_A$-term can be strong enough to justify Eq.\ \eqref{Qtau3}, see Ref.~\cite{we2017} for an estimate. This paper focuses on this case (absence of cooperons).

In the next sections of the Appendix, we derive the action for fluctuations $W$ using the action Eq.\ \eqref{appSmain}. We use the same saddle-point as the main text, given by Eqs.~\eqref{QLsaddle},\eqref{thetaEq}. Note that this is not an accurate saddle-point of the action Eq.\ \eqref{appSmain}. The exact saddle-point solution of this action has off-diagonal components in Nambu space in the magnetic region in the vicinity of the interface. This smoothens out the sharp kink of $Q_0(x)$ at the interface, producing an exponentially decaying tail in $\theta(x)$ in the normal region. This tail represents the weak proximity effect and gives rise to the exponentially small average Josephson current $\langle I\rangle\sim e^{-L/l_*}$. We neglect this tail since it leads to exponentially small contributions to the Josephson effect.

\section{Derivation of the action $S[W]$} \label{appS}
The constraints on $W$, defined by \eqref{Wparam} are
\begin{equation}
\{W,\Lambda\}=0,\quad
W= - \tau_1\sigma_1s_2W^T\tau_1\sigma_1s_2. \label{appWsym}
\end{equation}
The first ensures $Q^2=1$, the second comes from Eq.\ \eqref{appSelfconjugate}.

We start with rewriting the gradient term of the action Eq.\ \eqref{act1}. Using the notation $M = e^{-\frac12\tau_{\phi}(x) \tau_3\sigma_3\hat\theta(x)}\nabla e^{\frac12\tau_{\phi}(x) \tau_3\sigma_3\hat\theta(x)} = \tau_\varphi\tau_3\sigma_3\nabla{\theta}/2$ we have
\begin{multline}
\tr(\nabla Q)^2 = \tr \left( \nabla\Lambda e^{iW} +  M\Lambda e^{iW} - \Lambda e^{iW}M\right)^2
\\
= \tr \left\{(\nabla W)^2 + 2\left[M,\Lambda e^{iW}\right]\nabla\Lambda e^{iW}+ [M,\Lambda e^{iW}]^2\right\}
\\
=\tr(\nabla W)^2 - \tr[M,W]^2.
\end{multline}
The rest of the terms in Eq.\ \eqref{act1} are straight-forward to calculate making use of Eqs.\ \eqref{Wparam},\eqref{appWsym} and of the absence of magnetic terms in the superconducting lead.
In the end, we get the action
\begin{multline}\label{appSWS}
S_W^{(S)}= \frac{\pi\nu D}{8}\int dx \tr\biggl\{(\nabla W)^2 - \frac14[\tau_\phi\tau_3\sigma_3\nabla\hat\theta,W]^2 \\
 +\hat\varkappa^2\cos(\hat\theta_0-\hat\theta)W^2\biggr\}
\end{multline}
in the leads and
\begin{multline}\label{appSWM}
S_W^{(M)}= \frac{\pi\nu D}{8}\int dx \tr\left\{(\nabla W)^2 -\frac{1}{l_s^2}[\tau_3\mathbf{s},W]^2 \right.\\ \left. - \frac{1}{l_A^2}[\tau_3,W]^2  + \frac{2\hat\omega+2ih\sigma_3s_3}{D}W^2\right\}
\end{multline}
in the link.
The constraints of Eq.\ \eqref{appWsym} resolve into the following structure of $W$ in Nambu and particle-hole space
\begin{equation}
W = \begin{pmatrix}
d\sigma_1+d'\sigma_2 & c-ic' \\
c+ic' & -s_2(d^T\sigma_1+d'^T\sigma_2)s_2
\end{pmatrix}.
\end{equation}
Substituting this into Eqs.\ \eqref{appSWS},\eqref{appSWM} we find that actions for $c$ and $d$ separate: $S_W=S_d[d,d']+S_c[c,c']$.

We next consider S and M regions separately.

\subsection{Action in the magnetic region}\label{appSDN}

We start with calculating the action in the magnetic region M. Here, $U(x)$ equals $\sigma_1$ if $\omega<0$ and unity otherwise. For the diffuson action, after tracing over Nambu and particle-hole space, we arrive at
\begin{multline}
S^{(M)}_d = \frac{\pi\nu}{2} \int dx \tr \left\{
D (\nabla d)^2 + \frac{D}{l_s^2} [\mathbf s, d]^2 + 2\hat\omega d^2 \right.\\ \left. + (d\rightarrow d') - 2 hs_3 [d,d']\right\} .\label{Sd}
\end{multline}
In the case of magnetic impurities the exchange field $h$ is absent so that $d$ and $d'$ do not mix. We decompose $d$ into spin components via $d=\sum_{\alpha=0}^3 d_\alpha s_\alpha$ where $d_0$ is the spin singlet diffuson, and $d_{1,2,3}$ represent spin triplet modes. In an infinite wire with gaussian action Eq.\ \eqref{Sd} (with $h=0$), these diffusons have the following propagator
\begin{equation}
\langle d_\alpha^{ij}(q)d_\beta^{kl}(q')\rangle = \frac{\delta(q-q')\delta_{\alpha\beta}  \delta_{il}\delta_{kj}}{\nu\left(Dq^2 + \omega_i+ \omega_j + \left(1-\delta_{0\alpha}\right)\frac{8D}{l_s^2}\right)}, \label{diffusonM}
\end{equation}
where $i,j,k,l$ are replica indices and $q$ denotes momentum, i.e., the Fourier transform of $x$. The propagator for $d'$ is identical. The last term in the denominator of Eq.\ \eqref{diffusonM} shows that the singlet diffuson $d_0$ is not affected by magnetic impurities and has a thermal decay length of $k^{-1} = \sqrt{D/(\omega_i+\omega_j)}$, while the triplet is suppressed by magnetic impurities on the length scale $\sim l_s$. Since we have a long junction in the sense $l_s\ll L$, the triplet diffusons contribution to $Z$ can be neglected. Then, $d$ and the singlet diffuson are the same: $d=d_0s_0$.

In the case of a ferromagnetic junction the $l_s$-term is absent from the action, but $h\neq0$, producing coupling between $d$ and $d'$ via the last term in Eq.\ \eqref{Sd}. Rewriting this term as $\sim \tr[s_3,d]d'$ we see that $d_1$ is coupled to $d'_2$ (and $d_2$ to $d'_1$) while the spin components $d_0,d_3,d'_0,d'_3$ remain unaffected. The latter have the usual diffuson propagator
\begin{equation}
\langle d_\alpha^{ij}(q)d_\beta^{kl}(q')\rangle = \frac{\delta(q-q')\delta_{\alpha\beta}  \delta_{il}\delta_{kj}}{\nu\left(Dq^2 + \omega_i+ \omega_j \right)}, \label{diffusonF}
\end{equation}
with $\alpha=0,3$. The other half of the diffuson modes, $d_{1,2}$ get mixed with $d'_{2,1}$, forming decaying modes. Their propagator is proportional to $[Dq^2 + \omega_i+\omega_j \pm 2ih]^{-1}$. The $h$-term here gives rise to a complex $q$ with both exponential decay and oscillations on the scale $l_h=\sqrt{D/h}$. In long junctions, $l_h\ll L$ these modes can thus be neglected, just like the triplet diffusons in the magnetic impurity case.

Let us now consider the cooperon action in the magnetic region. Tracing out Nambu space, we get
 \begin{multline}
S^{(M)}_c = \frac{\pi\nu}{4} \int dx \tr \left\{
D (\nabla c)^2  + \frac{D}{l_s^2} \{\mathbf s, c\}^2 \right.
\\
\left.+ 2\left(\omega + \frac{2D}{l_A^2} + ihs_3\sigma_3\right)c^2 \right\} + (c\rightarrow c') .\label{Sc}
\end{multline}

We see that both the $l_s$ and the $l_A$ terms fully suppress all cooperons. Thus, they can be neglected when either magnetic impurities or orbital effects are present. The exchange term only suppresses half of the cooperon modes. This means that in a ferromagnet, some cooperons may be relevant. However, vector potentials are always present in ferromagnets, giving rise to the $l_A$ term. This means that in a ferromagnet there are two limits -- the relatively short link, $L\ll l_A$ where orbital effects can be neglected, so that half the cooperons are relevant, and the relatively long link $L\gg l_A$, where orbital effects are relevant and all cooperons are negligible.

Below we mainly focus on the case where all cooperons are suppressed. This corresponds to writing $W$ in the form Eq.\ \eqref{WEq}, with off-diagonal components absent.

\subsection{Diffuson action in the superconductor}\label{appSDS}
We next  take the diagonal part of $W$ in Nambu space and substitute it into Eq.\ \eqref{appSWS}. The result is
\begin{multline}\label{appeqSDS2}
S_d^{(S)} = \frac{\pi\nu D}{8}\int dx \tr\Biggl\{\begin{pmatrix}\nabla{d}\sigma_1&0\\0&-s_2\nabla{d}^T\sigma_1s_2\end{pmatrix}^2
\\
- \frac14\left[\begin{pmatrix}
0 & -e^{i\hat\phi} \\ e^{-i\hat\phi} & 0
\end{pmatrix}\sigma_3\nabla\hat\theta,\begin{pmatrix}d&0\\0&-s_2d^Ts_2\end{pmatrix}\sigma_1\right]^2
\\
+ \hat\varkappa^2\cos(\hat\theta_0-\hat\theta)\begin{pmatrix}d^2&0\\0&-s_2d^{T2}s_2\end{pmatrix}\Biggr\}+(d\rightarrow d').
\end{multline}
Using permutation and transposition properties of the trace, the first and third term simplify to
\begin{equation}
S_{1+3}= \frac{\pi\nu D}{2}\int dx\tr\left\{ (\nabla{d})^2+ \hat\varkappa^2\cos(\hat\theta_0-\hat\theta)d^2\right\},
\end{equation}
where the trace in $S_{1+3}$ applies to replica and spin spaces, while Nambu and particle-hole spaces have been traced out already. In the second term of Eq.\ \eqref{appeqSDS2} we expand the commutator, then trace out particle-hole space getting
\begin{multline}
S_2= \frac{\pi\nu D}{8}\int dx \tr\Biggl\{
\begin{pmatrix}
0 & -e^{i\hat\phi} \\ e^{-i\hat\phi} & 0
\end{pmatrix}^2(\nabla\hat\theta)^2\begin{pmatrix}d&0\\0&-s_2d^Ts_2\end{pmatrix}^2
\\
+ \left[\begin{pmatrix}
0 & -e^{i\hat\phi} \\ e^{-i\hat\phi} & 0
\end{pmatrix}\nabla\hat\theta\begin{pmatrix}d&0\\0&-s_2d^Ts_2\end{pmatrix}\right]^2\Biggr\}
\\
=\frac{\pi\nu D}{4}\int dx \tr\left\{
-(\nabla\hat\theta)^2d^2+e^{-i\hat\phi}\nabla\hat\theta de^{i\hat\phi}\nabla\hat\theta s_2d^Ts_2
\right\},
\end{multline}
where in the last line $\tr$ only acts in replica and spin spaces. Note that the ordering of the factors in the last term is important: $d$ is a matrix in replica space and generally does not commute with $\hat\phi$ and $\hat\theta$. The full action for $d$ in the superconducting lead becomes
\begin{multline}
S_{d}^{(\mathrm{S})}= \frac{\pi\nu D}{2}\int dx\tr\biggl\{ (\nabla{d})^2+ \biggl(\varkappa^2\cos(\theta_0-\theta)-\frac{(\nabla\theta)^2}{2}\biggr)d^2
\\
+ \frac12e^{-i\varphi}\nabla\theta de^{i\varphi}\nabla\theta s_2d^Ts_2 \biggr\}+(d\rightarrow d').
\end{multline}
The expression in the big round brackets is further simplified to $(\varkappa^2-(\nabla\theta)^2)$ with the help of the integral of motion $(\nabla\theta)^2 = 2\varkappa^2 (1-\cos(\theta-\theta_0))$ of the sine-Gordon equation \eqref{thetaEq}. Finally, in the case of magnetic impurities $d=d_0s_0$ and the $s_2$-matrices can be commuted out, producing Eq.\ \eqref{SdS}.

\section{Calculation of $\mathcal{F}$} \label{appF}
To calculate the eigenvalue product of Eq.\ \eqref{schr1} via the Gelfand-Yaglom theorem we turn the Schr\"odinger equation Eq.\ \eqref{schr1} into a set of matching equations. Finding the general form of $d_0(x)$ in each of the three regions and then matching wave functions and derivatives at $x=\pm L/2$ we arrive at a linear set of equations which has nontrivial solutions if $\lambda$, the effective energy in Eq.\ \eqref{schr1}, belongs to the spectrum. Thus, the determinant of this linear set of matching equations can be taken for $\mathcal{F}(\lambda)$. We will need to find $\mathcal{F}(0)$ so we put $\lambda=0$ and construct the wave functions. In this subsection we use a particular simple gauge, where the superconducting phase is zero in the left lead: $\hat{\phi}(x<-L/2)=0$ and $\hat{\phi}(x>L/2)=\hat{\varphi}$.
In the magnetic region, our hamiltonian has the form
\begin{align}
H_{ij}^{(\mathrm{M})} &= \pi\nu D\left(-\nabla^2 + k^2\right)\begin{pmatrix} 1 & 0 \\ 0 & 1\end{pmatrix},
\label{HN}
\\
k^2 &= \frac{\omega_i + \omega_j}{D}.
\end{align}
Thus $d$ consists of four plane waves with imaginary momenta $\pm ik$.
\begin{equation}\label{appSolM}
\begin{pmatrix}d_{0,ij} \\ d_{0,ji} \end{pmatrix}\left(|x|<\frac{L}2\right) =
\begin{pmatrix}\gamma_+ e^{kx}+\gamma_- e^{-kx} \\ \overline\gamma_+ e^{kx}+\overline\gamma_- e^{-kx} \end{pmatrix}.
\end{equation}
In the right superconducting lead, there are two eigenmodes decaying at $x\to+\infty$. The hamiltonian Eq.\ \eqref{HS} commutes with the isospin projection matrix
\begin{equation}
\begin{pmatrix} 0 & e^{i\delta\varphi_{ij}}\\
e^{-i\delta\varphi_{ij}} & 0\end{pmatrix},
\end{equation}
where $\delta\varphi_{ij}=\varphi_i-\varphi_j$. Therefore, the two decaying solutions of Eq.\ \eqref{schr1} have different isospin structure corresponding to the $\pm1$ eigenvalues of
the above operator in the corresponding superconducting lead. The wave function in the right lead is thus the linear combination:
\begin{equation}
\begin{pmatrix}d_{0,ij} \\ d_{0,ji} \end{pmatrix}\left(x>\frac{L}2\right) = \begin{pmatrix} \alpha_-\psi_-(x) + \alpha_+\psi_+(x)\\
\left(\alpha_-\psi_-(x) - \alpha_+\psi_+(x)\right)e^{-i\delta\varphi_{ij}}\label{appPsiR}
\end{pmatrix},
\end{equation}
where the functions  $\psi_\pm(x)$ are decaying solutions of the Schr\"odinger equation
\begin{equation}
-\nabla^2 \psi_\pm + U_\pm \psi_\pm =0, \label{appEqPsi}
\end{equation}
with potential
\begin{equation}
U_\pm = -\frac12\left((\nabla\theta_i)^2+(\nabla\theta_j)^2\pm\nabla{\theta}_i\nabla{\theta}_j - \varkappa_i^2 - \varkappa_j^2\right).
\end{equation}
In the left lead the general solution is the same as in the right lead, Eq.\ \eqref{appPsiR}, up to the mirror transform $x\rightarrow -x$ and adjustment of superconducting phase, which we gauged to zero in the left lead:
\begin{equation}
\begin{pmatrix}d_{0,ij} \\ d_{0,ji} \end{pmatrix}\left(x<-\frac{L}2\right) = \begin{pmatrix} \beta_-\psi_-(-x) + \beta_+\psi_+(-x)\\
\beta_-\psi_-(-x) - \beta_+\psi_+(-x)
\end{pmatrix}.\label{appPsiL}
\end{equation}
There are eight matching equations:
\begin{align}
\alpha_-\psi_- + \alpha_+\psi_+ &= \gamma_+e^{kL} + \gamma_-e^{-kL},\label{appMatch1}
\\
(\alpha_-\psi_- - \alpha_+\psi_+)e^{-i\delta\varphi_{ij}} &= \overline\gamma_+e^{kL} + \overline\gamma_-e^{-kL},\label{appMatch2}
\\
\alpha_-\nabla\psi_- + \alpha_+\nabla\psi_+ &= k(\gamma_+e^{kL} - \gamma_-e^{-kL}),\label{appMatch3}
\\
(\alpha_-\nabla\psi_- - \alpha_+\nabla\psi_+)e^{-i\delta\varphi_{ij}} &= k(\overline\gamma_+e^{kL} - \overline\gamma_-e^{-kL}),\label{appMatch4}
\\
\beta_-\psi_- + \beta_+\psi_+ &= \gamma_+ + \gamma_-,\label{appMatch5}
\\
\beta_-\psi_- - \beta_+\psi_+ &= \overline\gamma_+ + \overline\gamma_-,\label{appMatch6}
\\
\beta_-\nabla\psi_- + \beta_+\nabla\psi_+ &= -k(\gamma_+ - \gamma_-),\label{appMatch7}
\\
\beta_-\nabla\psi_- - \beta_+\nabla\psi_+ &= -k(\overline\gamma_+ - \overline\gamma_-),\label{appMatch8}
\end{align}
where $\psi_\pm,\nabla\psi_\pm$ are taken at $x=L/2$.
Equations \eqref{appMatch1} and \eqref{appMatch2} match wave functions at the right interface, $x=L/2$, Eqs.\ \eqref{appMatch3} and \eqref{appMatch4} match wave function gradients at $x=L/2$, the remaining Eqs.\ \eqref{appMatch5}-\eqref{appMatch8} do the same at the other interface, $x=-L/2$. The eight linear equations \eqref{appMatch1}-\eqref{appMatch8} involve eight variables $\alpha_\mp,\beta_\mp,\gamma_\mp,\overline{\gamma}_\mp$. Its determinant equals $\mathcal{F}_{ij}(0)$ by our definition of the function $\mathcal{F}(\lambda)$. After some algebra we get, up to $\varphi$-independent factors,
\begin{gather}
\mathcal{F}_{ij}(0) = 1  - \frac{e^{2kL}(\zeta_--\zeta_+)^2}{\left[e^{2kL}-\zeta_-\zeta_+\right]^2}\cos^2\frac{\delta\varphi_{ij}}2, \label{appF}
\\
\zeta_\mp = \left.\frac{k + \nabla \ln \psi_\mp(x)} {k - \nabla \ln \psi_\mp(x)}\right|_{x=L/2}.
\end{gather}
Only the logarithmic derivative of $\psi_\pm$ enters $\mathcal{F}_{ij}(0)$. Thus, it makes sense to introduce $q_\pm(x)=-\nabla\ln\psi_\pm(x)$. Substituting this into the Schr\"odinger equation, Eq.\ \eqref{appEqPsi}, we obtain Eq.\ \eqref{eqq}, and the expression Eq.\ \eqref{appF} turns into Eqs.\ \eqref{Fij} and \eqref{eqf}.

The function $\mathcal{F}_{ij}(-\infty)$ is obtained similarly and is also captured by the formula of the form Eqs.\ \eqref{Fij} and \eqref{eqf}. However, the effective momenta $k$ and $q_\pm$ must now be calculated at nonzero $\lambda$. At $\lambda\to-\infty$ the scaling is $k,q_\pm\sim\sqrt{-\lambda}$ with positive $k$. This means that $f$, Eq.\eqref{eqf} is dominated by the hyperbolic functions in its denominator, and scales as $e^{-\#\sqrt{\lambda}L}$, so that $\mathcal{F}_{ij}(\lambda\to-\infty)=1$.

\section{Calculation of $\left< I I \right>$ in various limits} \label{appII}
In this section we omit indices in $\delta\varphi_{ij}$ for brevity.

\subsection{High temperature/Long junction limit $T\gg E_\mathrm{Th}$}\label{appII1}
When the length $L$ exceeds the thermal length $\sqrt{D/T}$, the action $kL$ becomes large and $f\sim e^{-kL}\ll1$ so that
\begin{equation}
\langle I(\varphi_1)I(\varphi_1)\rangle = 2e^2T^2\cos\delta\varphi\sum\limits_{\omega_{1,2}}f^2. \label{appCorr}
\end{equation}
Since $kL\gg1$, we may neglect all terms except the first with $\omega_{1,2}=\pi T$. At $\omega_1=\omega_2$ Eq.\ \eqref{appEqPsi} becomes that of a particle in a $U_0/\cosh^2{\varkappa x}$ potential well at bound state energy and is solved exactly by $\psi_- = \nabla\theta$ and $\psi_+ = \nabla^2\theta$. We then find at the boundary $x=L/2$
\begin{align}
q_+ &=  2\sqrt{\frac{\omega}D}\left[1+\sqrt{1+\frac{\Delta^2}{\omega^2}}\right]^{-1/2},
\\
q_- &=  \sqrt{\frac\omega D}\left[1+\sqrt{1+\frac{\Delta^2}{\omega^2}}\right]^{1/2}.
\end{align}
Substituting this into $f$ and Eq.\ \eqref{appCorr} and keeping only the first term of the sum we obtain the result Eq.\ \eqref{eqIc1}.

\subsection{High temperature/low gap limit $T\gg \Delta$}\label{appII2}
Another situation where $q_\mp$ can be calculated analytically is the limit $\Delta\ll\omega_{1,2}$. In this case the superconducting terms in the potential $U_\pm$ can be treated perturbatively:
\begin{multline}
U_\mp = \frac12\left(\varkappa_1^2+\varkappa_2^2-(\nabla\theta_1)^2-(\nabla\theta_2)^2\pm \nabla\theta_1\nabla\theta_2\right) = \\ =k^2 + \delta U_\mp, \label{appUmp}
\end{multline}
with $\delta U_\mp\ll1$:
\begin{multline}
\delta U_\mp = \frac{\Delta^2}{D\omega_1}\left[\frac12-e^{-2k_1\left(|x|-\frac{L}{2}\right)}\right]
\\ +\frac{\Delta^2}{D\omega_2}\left[\frac12-e^{-2k_2\left(|x|-\frac{L}{2}\right)}\right]
\\
\pm \frac{\Delta^2}{D\sqrt{\omega_1\omega_2}}e^{-(k_1+k_2)\left(|x|-\frac{L}{2}\right)}+O\left(\frac{\Delta^4}{\omega^4}\right), \label{appdU}
\end{multline}
where we abbreviated $k_{i} = \sqrt{2\omega_i/D}$.

We look for a solution of Eq.\ \eqref{eqq} using the ansatz
\begin{gather}
q(x)=k+\delta q(x), \\
\nabla{\delta q} - 2k\delta q - \delta q^2+\delta U=0. \label{appeqs}
\end{gather}
Note that this ansatz and Eq.\ \eqref{appeqs} resemble the quasiclassical expansion procedure. Indeed, in the limit $\omega\gg \Delta$ the potential $U_\pm$ satisfies the well-known quasiclassicality condition $\nabla{U}\ll U^{3/2}$. However, the quasiclassical wave function $\psi_{qc}(x)\sim U(x)^{-1/4}\exp[-\int^x \sqrt{U}dx]$ is invalid for the potential $U_\pm$! This is nontrivial and deserves a detailed explanation. In a conventional quasiclassical limit $\hbar\to0$, effectively rescaling length $x\rightarrow y\hbar$, and producing a hierarchy of derivatives, $d^n/dy^n\propto \hbar^n$, with each additional $\partial_y$ producing an extra power in the small parameter $\hbar$. This hierarchy validates the quasiclassical expression $\psi_{qc}(x)$ and all further terms of the expansion in powers of $\hbar$. In such a quasiclassical regime, the first term in Eq.\ \eqref{appeqs} would be neglected (compared to the second term) as it contains a derivative. However, Eq.\ \eqref{eqq} in the $\Delta/\omega\to0$ limit is not truly quasiclassical and the $\nabla{\delta q}$ term must be kept. Considering derivatives of $U_\mp$ of Eq.\ \eqref{appUmp} we find $\nabla U \sim k U \Delta^2/\omega^2$ which indeed invokes the small parameter $\Delta/\omega$. However, further derivatives obey $\partial_x^n U \sim k^n U\Delta^2/\omega^2 $ -- no additional powers of $\Delta/\omega$ are generated! Thus, the problem at $T\gg\Delta$ is not truly quasiclassical, although the basic prerequisite $\nabla{U}\ll U^{3/2}$ is met (this inequality is sometimes mistaken for the criterion for the full quasiclassical expansion to work). In fact, the latter condition only justifies the main order quasiclassical approximation, i.e., $\psi(x) \sim \exp[-\int^x \sqrt{U}dx]$. Employing the quasiclassical $U^{-1/4}$ preexponent, (i.e., using the next term of the quasiclassical approximation) is wrong and leads to a wrong result.

Returning to Eq.\ \eqref{appeqs}, we assume $\delta q\ll1$ and neglect $\delta q^2$. Correspondingly we will only keep the lowest order term in $\delta U$ (i.e., all the three explicit terms in the right-hand side of Eq.\ \eqref{appdU}). We get
\begin{gather}
-\nabla{\delta q} + 2k\delta q-\delta U_\mp=0,
\\
\delta q(x) = \int\limits_x^{+\infty}\delta U_\mp(x')e^{2k(x-x')}dx'.
\end{gather}
This is the only solution satisfying $\delta q(x\to+\infty)\to0$. The function $\delta q(x)$ is maximal at $x=L/2$ (i.e., at the edge of the superconductor), where it equals
\begin{multline}
\delta q_\pm\left(\frac{L}2\right) = \int\limits_{0}^{+\infty}\delta U_\pm\left(x+\frac{L}{2}\right)e^{-2kx}dx
\\
=\frac{k\Delta^2}{4\omega_1\omega_2} - \frac{\Delta^2}{2\omega_1D(k+k_1)} - \frac{\Delta^2}{2\omega_2D(k+k_2)}
\\
\mp \frac{\Delta^2}{\sqrt{\omega_1\omega_2}D(2k+k_1+k_2)}. \label{appdots}
\end{multline}
This is much smaller than unity justifying the neglection of $\delta q^2$ in Eq.\ \eqref{appeqs}. Substituting this into Eq.\eqref{Fij} and using $q_\mp\approx k$ in the denominator, we get, up to $\delta\varphi$-independent factors,
\begin{equation}
\mathcal{F}_{12} \approx 1-\frac{e^{-2kL}}{4k^2}[q_+-q_-]^2\cos^2\frac{\delta\varphi}{2} + O\left(\frac{\Delta^6}{\omega^6}\right),
\end{equation}
and the current-current correlator becomes
\begin{multline}
\langle I_{1}I_2\rangle=2e^2T^2\cos\delta\varphi\sum\limits_{\omega_{1,2}}\frac{e^{-2kL}}{4k^2}[q_+-q_-]^2
\\
= \frac{e^2\Delta^4e^{-\sqrt{\frac{8\pi T}{E_\mathrm{Th}}}}}{32\pi^4T^2}\alpha^2\left(\frac{T}{E_\mathrm{Th}}\right),
\end{multline}
where
\begin{multline}
\alpha^2(\gamma) = \sum\limits_{n_{1,2}=0}^\infty e^{\sqrt{8\pi\gamma}-\sqrt{8\pi\gamma(n_1+n_2+1)}}
\\
\times\frac{16(\sqrt{2n_1+1}+\sqrt{2n_2+1}+2\sqrt{n_1+n_2+1})^{-2}}{(2n_1+1)(2n_2+1)(n_1+n_2+1)}.
\end{multline}
is a monotonous function of $\gamma$ quickly decreasing from $\alpha(0)=1.146$ to $\alpha(+\infty)=1$, in which limit only the first term survives exponential suppression.

\subsection{Low temperature/long junction limit $\Delta\gg E_\mathrm{Th},T$}\label{appII3}
Owing to the $e^{-kL}$ exponent, the summation over $\omega_{1,2}$ is dominated by frequencies of the order or less than $E_\mathrm{Th}$. Thus, if $\Delta$ is the largest parameter, then the summation happens over energies much smaller than $\Delta$ and we may use the small parameter $\omega/\Delta$. We have
\begin{gather}
\zeta_\mp = \mp 1 + O\left(\sqrt{\frac{\omega}{\Delta}}\right),
\\
\mathcal{F}_{12}(0) = \cosh 2kL - \cos\delta\varphi,
\\
\langle I(\varphi_1)I(\varphi_2)\rangle = 4e^2T^2\sum\limits_{\omega_{1,2}}\frac{\partial^2}{\partial\delta\varphi^2}\ln \left[\cosh 2kL - \cos\delta\varphi\right].
\end{gather}
Since $kL=\sqrt{(\omega_1+\omega_2)/E_\mathrm{Th}}$ only depends on the sum of frequencies, the double summation over Matsubara frequencies $\omega_1,\omega_2$ reduces to a single sum over $\omega_1+\omega_2 = 2m\pi T$ with natural $m$:
\begin{multline}
\langle I(\varphi_1)I(\varphi_2)\rangle = \\
=4e^2T^2\sum\limits_{m=1}^\infty m\frac{\partial^2}{\partial\delta\varphi^2}\ln \left[\cosh \sqrt{\frac{8m\pi T}{E_\mathrm{Th}}} - \cos\delta\varphi\right].
\end{multline}
In the subcase $E_\mathrm{Th}\ll T\ll\Delta$ the first term, $m=1$, dominates and we reproduce the long-junction result Table~I, case 1, see also Eq.\ \eqref{eqIc1}. In the opposite subcase $T\ll E_\mathrm{Th} \ll\Delta$ which corresponds to a long junction (in the sense $\sqrt{D/\Delta}\ll L$) at very low temperatures, the sum over frequencies can be replaced by integration, and we find
\begin{multline}
\langle I(\varphi_1)I(\varphi_2)\rangle =
\frac{e^2E^2_\mathrm{Th}}{8\pi^2}\frac{\partial}{\partial\delta\varphi}\sin\delta\varphi\int\limits_0^\infty  \frac{y^3dy}{\cosh y - \cos\delta\varphi}
\\
=\frac{3e^2E^2_\mathrm{Th}}{2\pi^2}\Re\mathrm{Li}_3(e^{i\delta\varphi}) =\frac{3e^2E^2_\mathrm{Th}}{2\pi^2}\sum\limits_{n=1}^\infty \frac{\cos (n\delta\varphi)}{n^3}.
\end{multline}
The presence of higher harmonics in $\langle I(\varphi_1)I(\varphi_2)\rangle$ indicates the presence of higher harmonics in $I(\varphi)$ as well.

\subsection{Low temperature/Short junction limit $E_\mathrm{Th}\gtrsim \Delta \gtrsim T$}
The only corner of our $E_\mathrm{Th}/\Delta,T/\Delta$ diagram, see Fig.~\ref{FigDiagram}, bottom right, that is not covered by the previous three limits is $E_\mathrm{Th}\gg \Delta \gg T$, which corresponds to a short junction (with respect to the superconducting coherence length) at low temperature (with respect to the gap). In this case, all harmonics are present, and the Matsubara integral converges over energies of the order of $\Delta$. Thus, there is no small parameter and Eq.\ \eqref{eqq} needs to be solved numerically. We have numerically found $I_t$, as well as the first five harmonics $J_{1,\dots5}$ at $L=0$ for different temperatures, with the results presented in Fig.~\ref{figShort}.

While $I_t,J_n$ cannot be calculated analytically, the derivative $d I_t^2/dT$ can. This is because it is dominated by small frequencies of the order of temperature. At low frequencies we have
\begin{equation}
f = -1+\frac{2(\omega_1+\omega_2)}{\Delta} + O\left(\frac{\omega^2}{\Delta^2}\right),
\end{equation}
so that
\begin{multline}
I_t^2=2e^2T^2\sum\limits_{\omega_1,\omega_2}\frac{f^2}{1-f^2}
\\
\approx2e^2\Delta T^2\sum\limits_{n_1,n_2=0}^\infty\frac{1}{4\pi T(n_1+n_2+1)}
=\frac{e^2\Delta T}{2\pi}\sum\limits_{n=1}^\infty 1.
\end{multline}
This obviously diverges, but we are only interested in the derivative $\partial/\partial T$ which is finite and governed by low-energy behavior. Replacing the summand $1$ with some function $g(mT)$ that regularizes the sum at $n\to\infty$ and applying the Euler-Maclaurin formula, we get
\begin{equation}
\sum\limits_{n=1}^\infty Tg(Tn) \approx  \int\limits_T^\infty g(y)dy+\frac{Tg(0)}{2},
\end{equation}
hence
\begin{equation}
\frac{\partial}{\partial T}\sum\limits_{n=1}^\infty Tg(Tn) \approx -\frac{1}{2}g(0),
\end{equation}
so that
\begin{equation}
\frac{\partial I_t^2}{\partial T} = -\frac{e^2\Delta}{4\pi},
\end{equation}
in agreement with our numerical findings.


\begin{thebibliography}{99}
\bibitem{NoteDiffusons}
Strictly speaking, $d$ and $d'$ represent diffusons in the conventional sense of soft modes with counterpropagating electrons only in the magnetic region. In the superconducting leads, $d$ and $d'$ are linear combinations of conventional cooperons and diffusons.
\bibitem{commentDet}
We omit here the constant factor $\pi^{\mathrm{dim} H}$, since it goes to unity in the replica limit $n_{1,2}\to0$ and is thus irrelevant.
\bibitem{NoteTRS}
The spinless TRS $H=H^*$ holds not only for uniform exchange fields, but for any coplanar field $\mathbf{h}(\mathbf{r})$: the exchange term $h_x(\mathbf{r})s_x+h_z(\mathbf{r})s_z$ is purely real.
\end{thebibliography}
\end{document}